\documentclass[final,5p,times,twocolumn]{elsarticle}

\usepackage[english]{babel}
\usepackage[T1]{fontenc}
\usepackage{etoolbox}
\makeatletter
\patchcmd{\ps@pprintTitle}{\footnotesize\itshape
      Preprint submitted to \ifx\@journal\@empty Elsevier
      \else\@journal\fi\hfill\today}{\scriptsize{Preprint submitted to ASOC \hfill \today}}{}{}

\usepackage{float}
\usepackage{natbib}
\usepackage{graphicx}
\usepackage[lined,  boxed,  linesnumbered,  ruled]{algorithm2e}
\graphicspath{{Figures/}}
\usepackage{amssymb, amsfonts, amsmath, graphicx, epstopdf, bbding, lipsum, multirow, makecell, picinpar, enumerate}
\usepackage[table]{xcolor}
\usepackage{colortbl}
\usepackage{xspace}

\usepackage{subfigure} 

\makeatother
\usepackage{comment}
\usepackage{amsmath}

\usepackage{amsmath,amssymb,amsfonts}
\usepackage{algorithmic}
\usepackage[]{graphicx}
\graphicspath{{Figures/}}
\usepackage{lipsum}
\usepackage{balance}
\usepackage{textcomp}
\usepackage{xcolor}
\usepackage{multirow}
\usepackage{color}
\usepackage{lscape}
\usepackage{longtable,booktabs}
\usepackage[hidelinks]{hyperref}

\begin{document}

\begin{frontmatter}

\title{A multimodal Transformer for InSAR-based ground deformation forecasting with cross-site generalization across Europe}

\author[add1,add2]{Wendong~Yao}
\ead{wendong.yao@ucdconnect.ie}
\author[add1,add2]{Binhua~Huang}
\ead{binhua.huang@ucdconnect.ie}
\author[add1,add2]{Soumyabrata~Dev\corref{mycorrespondingauthor}}

\address[add1]{The ADAPT SFI Research Centre, Dublin, Ireland}
\address[add2]{School of Computer Science, University College Dublin, Belfield, Dublin, Ireland}
\cortext[mycorrespondingauthor]{Corresponding author. Tel.: + 353 1896 1797.}
\ead{soumyabrata.dev@ucd.ie}

\begin{abstract}
Near-real-time regional-scale monitoring of ground deformation is increasingly required to
support urban planning, critical infrastructure management, and natural hazard mitigation.
While Interferometric Synthetic Aperture Radar (InSAR) and continental-scale services such as
the European Ground Motion Service (EGMS) provide dense observations of past motion, predicting
the next observation remains challenging due to the superposition of long-term trends, seasonal
cycles, and occasional abrupt discontinuities (e.g., co-seismic steps), together with strong
spatial heterogeneity.

In this study we propose a multimodal patch-based Transformer for single-step, fixed-interval
next-epoch nowcasting of displacement maps from EGMS time series (resampled to a 64×64 grid over
100 km × 100 km tiles). The model ingests recent displacement snapshots together with (i) static
kinematic indicators (mean velocity, acceleration, seasonal amplitude) computed in a leakage-safe
manner from the training window only, and (ii) harmonic day-of-year encodings. On the eastern
Ireland tile (E32N34), the STGCN is strongest in the displacement-only setting, whereas the
multimodal Transformer clearly outperforms CNN–LSTM, CNN–LSTM+Attn, and multimodal STGCN when
all models receive the same multimodal inputs, achieving RMSE = 0.90 mm and R² = 0.97 on the test
set with the best threshold accuracies.

We further assess transferability by training a single model on E32N34 and applying it, without
fine-tuning, to five unseen European tiles spanning continuous subsidence, periodic motion, and
co-seismic deformation. Across all six tiles the model maintains R$^2\geq$  0.93, with RMSE ranging
from 0.7 to 3.2 mm and 71–87\% of pixels within 1 mm. For co-seismic tiles, the high overall
accuracy reflects robustness to strong time-series discontinuities and rapid post-event
re-alignment once the step is present in the input history, rather than anticipation of earthquake
occurrence. These results indicate that multimodal Transformers can serve as accurate local
nowcasters and transferable priors for EGMS-based deformation monitoring with limited local
training data.

\end{abstract}


\end{frontmatter}

\section{Introduction}
\label{sec:introduction}

Ground deformation driven by groundwater extraction, underground construction,
hydrocarbon production and tectonic processes poses a persistent threat to
buildings, transportation networks and critical infrastructure worldwide
\citep{GroundWater,UrbanDef1,UrbanDef2,UrbanDef3}.
Interferometric Synthetic Aperture Radar (InSAR) has become a key technology
for mapping such deformation with millimetre accuracy over large areas
\citep{OSMANOGLU201690,SAR1,SAR2}, and continental-scale services such as the
European Ground Motion Service (EGMS) now deliver harmonised ground-motion
products across Europe \citep{EGMSDescript,EGMS}.
However, these products are intrinsically retrospective: they describe how the
ground has moved so far, but not how it is likely to move at the next
acquisitions.
In many operational settings, even short-term information at the cadence of
new satellite passes (on the order of a few days for Sentinel-1 over Europe)
can help operators track the evolution of deformation signals and prioritise
sites for further investigation, complementing more traditional geodetic and
engineering assessments rather than replacing them.

Within Europe, these capabilities have recently been consolidated in the
European Ground Motion Service (EGMS), an operational Copernicus Land
Monitoring Service product based on Sentinel-1 SAR data \citep{EGMSDescript,EGMS}.
EGMS delivers homogenised, quality-controlled ground motion information over the
European continent, including line-of-sight displacement time series and
summary statistics such as mean velocity, linear acceleration and
seasonal-motion parameters on a regular grid.
This open and systematic service removes the need for bespoke InSAR processing
for many applications and offers, for the first time, continental-scale time
series of ground motion suitable for statistical analysis and machine-learning
modelling \citep{CROSETTOEGMS}.

Most current uses of EGMS and other InSAR products remain retrospective:
they focus on mapping and attribution of observed deformation, for example
by identifying subsiding urban districts, unstable slopes or regions of
aquifer compaction \citep{AZADNEJAD2020101950,environsciproc2023028015}.
Only a limited number of studies have attempted to \emph{forecast} future
ground motion using data-driven models, despite the potential value of
short-term predictive capability for monitoring and maintenance workflows
\citep{KUZUEGMS}.
Recent work has begun to explore deep learning approaches for InSAR time
series, for instance using convolutional and recurrent neural networks to
detect deformation patterns or to model local displacement histories at
specific sites \citep{KUZUEGMS}.
However, these applications are typically restricted to one or a few locations,
treat each pixel independently, or rely on relatively simple network
architectures.

Spatio-temporal deep learning models designed for structured data provide
a natural way to exploit both the temporal dynamics and spatial
correlation of InSAR ground-motion fields.
Spatio-temporal graph convolutional networks (STGCNs) \citep{yu2017spatio},
originally developed for traffic forecasting, combine temporal convolutions
with graph convolutions defined on a fixed adjacency structure and have
proven effective for learning complex propagation patterns in networked
systems.
Self-attention and Transformer architectures, first proposed for natural
language processing, have subsequently been adapted for time-series forecasting
and for modelling spatial--temporal processes, allowing models to learn
long-range and non-local dependencies without an explicit graph.

At the same time, most existing forecasting studies in ground deformation
still adopt a uni-modal perspective, using only the displacement time
series as input \citep{KUZUEGMS}.
In contrast, operational products such as EGMS expose a richer set of
information \citep{EGMS}: in addition to per-epoch displacement, each
grid cell is associated with static motion descriptors (e.g. mean
velocity, linear acceleration, seasonal amplitude) summarising its
long-term behaviour, and the acquisition dates themselves encode clear
seasonal cycles.
Incorporating this multi-modal information into spatio-temporal deep
learning models has the potential to improve forecasting skill, but this
opportunity has not yet been systematically explored.

In our previous work we proposed a convolutional neural
network--long short-term memory (CNN--LSTM) architecture for short-term
InSAR ground deformation forecasting over eastern Ireland, using only
sequences of displacement maps as input \citep{yao2025deeplearningapproachspatiotemporal}.
While this uni-modal approach achieved promising skill, particularly when
combined with an STGCN baseline on the same grid, its accuracy remained
limited in regions with strong seasonal variability or low signal-to-noise
ratio, and it did not explicitly exploit the static descriptors already
available from the EGMS processing chain.

Also, this manuscript is related to our earlier study on EGMS/InSAR deformation forecasting~\citep{yao2025multimodalspatiotemporaltransformerhighresolution},
which investigated window-based forecasting models that are primarily trained and evaluated in a site-specific manner.
In contrast, the focus here is on \emph{transferable} single-step forecasting:
we introduce a multimodal formulation that leverages EGMS-derived static deformation indicators and harmonic time encodings,
and we develop a patch-based Transformer that is evaluated under cross-site regimes, including zero-shot transfer across multiple European tiles.
Accordingly, the model design choices, experimental protocols (cross-site and leakage-safe feature construction), and conclusions differ from~\citep{yao2025multimodalspatiotemporaltransformerhighresolution}.

The present study addresses these gaps by moving from a uni-modal to a
multi-modal representation of EGMS ground motion and by systematically
evaluating both graph-based and attention-based spatio-temporal deep
learning architectures.
We construct a unified grid-based data set for a
$100~\mathrm{km} \times 100~\mathrm{km}$ EGMS tile covering eastern Ireland, in
which each sample consists of: (i) a sequence of displacement maps;
(ii) three static ground-motion indicator maps derived from the EGMS
Level~3 product (mean velocity, acceleration and seasonal amplitude); and
(iii) sinusoidal encodings of the acquisition day-of-year.
On this multi-modal input we train and compare four architectures: a
convolutional--recurrent CNN--LSTM, its attention-augmented variant,
a multi-modal STGCN and a newly designed spatio-temporal Transformer
that operates on multi-channel image patches with global self-attention
and residual next-step prediction.

In this work we focus on \emph{single-step}, \emph{fixed-interval}
forecasting at the native EGMS/Sentinel-1 cadence, i.e. predicting the
displacement map at the next available acquisition given a history of past
observations and auxiliary variables.
Such next-epoch forecasts are best viewed as a form of deformation
\emph{nowcasting} and short-term trend projection, intended to support
routine monitoring, screening and infrastructure management rather than
stand-alone, long-horizon early-warning systems.

Our main contributions are threefold:

\begin{itemize}
  \item We construct a multimodal learning framework that combines recent
  displacement maps with static auxiliary layers and temporal encodings,
  enabling the model to jointly exploit spatial susceptibility, seasonal
  forcing and local deformation history. We show that this multimodal design
  substantially improves single-step prediction accuracy over using
  displacement maps alone.

  \item We design a dedicated Transformer architecture tailored to
  InSAR-based deformation forecasting and benchmark it against convolutional
  sequence models (CNN--LSTM, CNN--LSTM with attention) and a state-of-the-art
  spatio-temporal graph convolutional network (STGCN). On a tile in eastern
  Ireland, the proposed multimodal Transformer achieves the best overall
  single-step performance across RMSE, MAE, $R^2$ and multiple
  error-threshold accuracies.

  \item Beyond local experiments, we systematically evaluate the spatial
  transferability of the multimodal Transformer. A model trained only on a
  single EGMS tile (E32N34) generalises well to five additional tiles
  representing continuous subsidence, periodic motion, co-seismic
  displacement and quasi-stable conditions, maintaining $R^2 \ge 0.93$ and
  high sub-millimetre accuracy. This demonstrates the potential of
  Transformer-based models as transferable priors for large-scale,
  EGMS-driven deformation nowcasting in regions with limited local
  training data.
\end{itemize}

\section{Data and study areas}
\label{sec:data}

\subsection{European Ground Motion Service data}

Our analysis is based on the Level~3 (L3) products of the European Ground
Motion Service (EGMS), part of the Copernicus Land Monitoring Service.
EGMS provides harmonized ground-motion information over the European
continent derived from multi-temporal interferometric processing of
Sentinel-1 SAR images \cite{EGMSDescript,EGMS}. The L3 product is a
regularly gridded data set in which each grid cell aggregates one or
more persistent or distributed scatterers and stores a displacement time
series along with several summary statistics and quality indicators.
For each grid cell, the L3 Up component used in this study contains:
\begin{itemize}
  \item the time series of vertical displacement with respect to a
        reference epoch;
  \item the mean velocity estimated from a linear trend fit to the
        time series;
  \item a linear acceleration term capturing long-term curvature;
  \item parameters describing the amplitude and phase of annual
        seasonal motion, derived from harmonic regression; and
  \item various uncertainty and quality metrics, including the standard
        deviation of residuals.
\end{itemize}

The underlying Sentinel-1 constellation provides a revisit time of
6~days over Europe in the ascending and descending orbits, enabling
dense temporal sampling of ground motion at a nominal spatial
resolution of approximately 20--30~m. The EGMS processing chain applies
consistent orbit corrections, atmospheric delay mitigation, geocoding
and rigorous quality control, producing a homogeneous long-term record
of surface deformation suitable for regional and continental-scale
applications. In this work we exploit this consistency to study both
local forecasting skill and cross-site generalization of our
multimodal Transformer across different deformation regimes in Europe.

For each 100~km~$\times$~100~km tile, the L3 Up product is delivered as
a comma-separated values (CSV) file named
\texttt{EGMS\_L3\_EXXNYY\_100km\_U\_2018\_2022\_1.csv}, where
\texttt{EXXNYY} encodes the tile identifier in the EGMS tiling grid. The
file contains all grid cells in the tile, with columns storing the
geographic coordinates, ground-motion summary statistics and a series of
displacement values for each Sentinel-1 acquisition between 2018 and
2022. We focus on this recent 5-year period to align with the
operational EGMS version and to avoid potential artefacts associated
with the transition from the Sentinel-1A-only to the Sentinel-1A/B
constellation.

\subsection{Study areas and deformation regimes}

To evaluate both local performance and cross-site generalization, we
select six EGMS tiles distributed across Europe,
chosen to represent a broad spectrum of ground-motion behaviours. Tiles
are identified using the standard EGMS naming convention, \textbf{EXXNYY},
where:
\begin{itemize}
  \item \textbf{XX} denotes the easting coordinate (in 100~km units,
        EPSG:3035) of the south-west corner of the tile's lower-left
        pixel; and
  \item \textbf{YY} denotes the corresponding northing coordinate.
\end{itemize}

The six tiles are grouped into three categories according to their
dominant deformation regime:

\begin{itemize}
  \item \textbf{Continuous and periodic deformation:}
        Tiles E32N34 and E32N35 are located on the east coast of Ireland
        and encompass the greater Dublin region and surrounding
        low-lying coastal plain. Previous work has shown that these
        tiles exhibit a rich mixture of slow subsidence bowls,
        localized uplift and pronounced seasonal signals related to
        hydrogeological forcing and anthropogenic loading. Tile E32N34
        serves as our primary \emph{source} tile for training the
        multimodal Transformer, while both E32N34 and E32N35 are used to
        assess forecasting skill in regions dominated by gradual,
        often quasi-periodic deformation.

  \item \textbf{Long-term subsidence:}
        Tiles E39N30 and E44N23 provide examples of more monotonic,
        long-term subsidence. These tiles cover urban and industrial
        areas underlain by compressible sediments and engineering
        structures where subsidence is driven by natural consolidation,
        groundwater extraction and surface loading. They are used to
        test whether a model trained on mixed-periodic behaviour can
        extrapolate persistent downward trends in previously unseen
        settings.

  \item \textbf{Abrupt co-seismic displacement:}
        Tiles E48N24 and E58N17 encompass regions that experienced
        strong co-seismic deformation during recent moderate-to-large
        earthquakes. These data are characterized by sharp spatial
        gradients and step-like temporal offsets superimposed on
        otherwise relatively stable background motion. They constitute
        the most challenging scenario for cross-site generalization,
        probing the ability of the model to reproduce sudden structural
        breaks in the time series.
\end{itemize}

Across all tiles, the EGMS L3 product provides a dense sampling of ground
motion over built-up areas, transport corridors, exposed rock and other
coherent scatterers. As is typical for C-band SAR, coverage is sparser
in vegetated and agricultural zones, but the gridded product retains
sufficient density to enable interpolation to a regular raster
representation suitable for convolutional and Transformer-based neural
networks.

\subsection{Pre-processing and feature construction}

To build a common input representation for all models and tiles, we
convert the irregular grid of EGMS L3 points into a stack of raster maps
at a spatial resolution of $64\times64$~pixels covering the full
100~km~$\times$~100~km extent of each tile. For every acquisition epoch,
the vertical displacement values are interpolated onto this grid using
linear scattered-data interpolation in the projected (easting,
northing) coordinate system, yielding a displacement cube of size
$T\times H\times W$ per tile.

We then define a chronological training window covering the first
$80\,\%$ of available epochs ($t=1,\dots,T_{\mathrm{train}}$) and a test
window comprising the remaining $20\,\%$ ($t=T_{\mathrm{train}}+1,\dots,T$).
All static deformation indicators used in this work are \emph{recomputed}
from the training window only, in order to avoid leakage of information
from future epochs. Concretely, for each grid cell we fit a simple
polynomial-plus-harmonic model to the training portion of the time
series and derive (i) a mean velocity, (ii) a quadratic acceleration
term and (iii) the amplitude of the dominant annual component. These
three statistics are then interpolated onto the same $64\times64$ grid
to form time-invariant ``static'' maps per tile. We do not use the
EGMS L3 summary parameters that are fitted over the full 2018--2022 time
series.

For a given tile, the resulting data cube has dimensions
$T \times C \times H \times W$, where $T$ is the number of acquisition
epochs (about 300 in our case), $H=W=64$, and $C$ denotes the number of
input channels. For the uni-modal setting we use only the displacement
channel ($C=1$). For the multimodal setting we construct a six-channel
representation ($C=6$) consisting of:
\begin{enumerate}
  \item the displacement map at each epoch;
  \item three static maps encoding mean velocity, acceleration and
        seasonal amplitude, all estimated from the training window only;
  \item two temporal encoding channels given by the sine and cosine of
        the normalised day-of-year of the acquisition date, which
        capture the annual cycle in a smooth, periodic form.
\end{enumerate}

Before model training, all channels are standardised.
Static channels are normalised per-channel using their spatial mean and
standard deviation. Displacement time series are standardised on a
per-pixel basis using statistics computed from the training portion of
the corresponding tile, and the same affine transformation is then
applied to all time steps of that tile. This prevents information
leakage from the test window into the training process while preserving
the relative amplitudes of seasonal and long-term motions.

Finally, we construct supervised learning samples by sliding a temporal
window of length $T_{\text{in}}$ over the normalised displacement cube.
For each window, the input consists of $T_{\text{in}}$ consecutive
multi-channel maps, and the target is the displacement map at the next
time step. In this paper we focus on single-step forecasting with
$T_{\text{in}}=16$ for the Transformer and a comparable historical
context for the other models. Sequence pairs are split in chronological
order into a training set (first 80\% of samples) and a validation/test
set (remaining 20\%), ensuring that all models are evaluated on strictly
unseen future data. For the cross-site generalisation experiments, the
network is trained only on tile E32N34 using its training window, and
then applied, without fine-tuning, to the test windows of the remaining
tiles using the same input representation and normalisation strategy.

\subsubsection{Resolution sensitivity analysis}
In this work we deliberately operate at a regional grid resolution of
$64\times64$ pixels over each $100\text{ km}\times100\text{ km}$ tile,
corresponding to a pixel spacing of approximately $1.56\text{ km}$.
This choice reflects a trade–off between spatial detail and the memory
requirements of training multimodal 3D spatio–temporal networks on
several hundred Sentinel–1 epochs. Increasing the grid to
$128\times128$ or $256\times256$ would lead to a $4\times$ or
$16\times$ increase in the number of spatial degrees of freedom per
time step, which is prohibitive for the multimodal Transformer under our
GPU budget.

Rather than targeting infrastructure–scale motion at the level of
individual buildings, we focus here on regional–scale forecasting of
ground deformation on a $\sim1.5\text{ km}$ grid, which is directly
compatible with tile–wise EGMS products and suitable for city– and
basin–scale risk screening.

To address the potential influence of spatial downsampling on forecast
skill, we conducted a small-scale sensitivity experiment on tile E32N34.
In the main experiments, all models operate on rasterised EGMS L3 inputs
resampled to a $64\times64$ grid (Section~\ref{sec:data}), corresponding
to a pixel spacing of approximately 1.5~km. Here we retrain a lighter
CNN--LSTM baseline, using the same displacement-only configuration and
training protocol as in Section~\ref{sec:experiments}, but at two
different spatial resolutions: $64\times64$ and $128\times128$.

Table~\ref{tab:res_sensitivity} summarises the results on the last
20\,\% of the time series for both settings. Increasing the resolution
from $64\times64$ to $128\times128$ yields a modest reduction in
$\mathrm{RMSE}$ from 1.96~mm to 1.79~mm and a slight increase in
$R^{2}$ from 0.87 to 0.88. Threshold accuracies are essentially
unchanged: $\mathrm{Acc@1mm}$ improves only from 60.0\,\% to 60.8\,\%,
and the tighter thresholds (0.5, 0.2 and 0.1~mm) differ by less than
0.5 percentage points.

\begin{table}[t]
    \small
    \setlength{\tabcolsep}{1mm}
    \centering
    \caption{Resolution sensitivity for a CNN--LSTM baseline on tile
    E32N34 (displacement-only, SmoothL1 loss). Metrics are computed on
    the last 20\,\% of epochs.}
    \label{tab:res_sensitivity}
    \begin{tabular}{lcccc}
        \toprule
        Resolution & RMSE [mm] & MAE [mm] & $R^2$ & Acc@1mm [\%] \\
        \midrule
        $64\times64$  & 1.96 & 1.16 & 0.87 & 60.02 \\
        $128\times128$ & 1.79 & 1.11 & 0.88 & 60.78 \\
        \bottomrule
    \end{tabular}
\end{table}

These findings indicate that, for the EGMS L3 products and deformation
patterns considered in this study, doubling the grid resolution does not
lead to substantial gains in single-step forecast performance. The
64$\times$64 representation therefore appears sufficient to capture the
dominant spatial structure of the displacement field, while providing a
favourable trade-off between fidelity and computational efficiency for
training the more demanding multimodal Transformer and STGCN models.

\section{Methodology}
\label{sec:methods}

\subsection{Problem formulation}

Let $\Omega \subset \mathbb{R}^2$ denote the spatial domain of an EGMS
tile, discretised on a regular $H \times W$ grid, and let
$t \in \{1,\dots,T\}$ index the Sentinel--1 acquisition epochs. For each
grid cell $(i,j) \in \{1,\dots,H\}\times\{1,\dots,W\}$ and time $t$, the
EGMS L3 Up product provides a vertical displacement
$d_t(i,j)$ in millimetres with respect to a reference epoch
(Section~\ref{sec:data}).

We cast short-term ground deformation forecasting as a supervised
spatio-temporal learning task. Given a sequence of $L$ historical
displacement maps
\begin{equation}
    \mathbf{D}_{t-L+1:t}
    = \{ d_{\tau}(i,j) \mid \tau = t-L+1,\dots,t;\; (i,j)\in\Omega \},
\end{equation}
together with time-invariant descriptors $\mathbf{S}(i,j)$ of the local
deformation regime and explicit temporal encodings
$\mathbf{h}_\tau$ of acquisition date, the goal is to predict the
displacement field at the next epoch,
\begin{equation}
    \hat{d}_{t+1}(i,j) = f_{\theta}\!\left(
        \mathbf{D}_{t-L+1:t},\,
        \mathbf{S},\,
        \mathbf{h}_{t-L+1:t+1}
    \right),
\end{equation}
where $f_{\theta}$ is a deep neural network with learnable parameters
$\theta$. Throughout this work we focus on single-step forecasting
($t+1$), but the formulation naturally extends to multi-step horizons.

As detailed in Section~\ref{sec:data}, we construct supervised samples by
sliding a fixed-length temporal window of size $L=T_{\text{in}}$ along
each EGMS time series. Each window yields an input tensor
$\mathbf{X}_n \in \mathbb{R}^{L \times C \times H \times W}$ and a
corresponding target
$\mathbf{Y}_n \in \mathbb{R}^{1 \times 1 \times H \times W}$, where $C$
is the number of input channels after fusing all modalities. We adopt a
chronological $80/20$ split in the temporal dimension: the first $80\%$
of epochs are used for training and the remaining $20\%$ for validation
and testing, thereby avoiding information leakage from future
acquisitions.

\begin{figure*}[t]
  \centering
  \includegraphics[width=\textwidth]{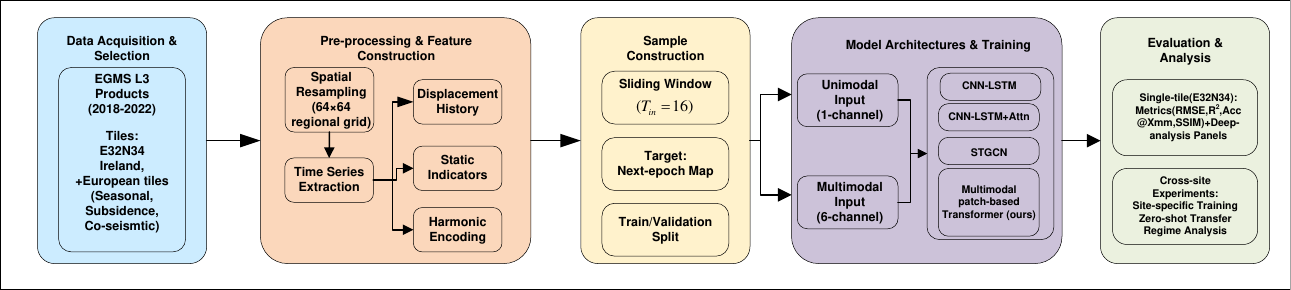}
  \caption{Overall workflow of the proposed EGMS-based multimodal deformation nowcasting framework, from EGMS L3 data acquisition through pre-processing and sample construction to model training and single-tile / cross-site evaluation.}
  \label{fig:flowdiagram}
\end{figure*}

Figure~\ref{fig:flowdiagram} shows overall workflow of our proposed EGMS-based multimodal deformation nowcasting framework.
\subsection{Multimodal input representation}
\label{subsec:input-repr}

\subsubsection{Displacement-only representation}

In our previous work~\citep{yao2025deeplearningapproachspatiotemporal}
we modelled EGMS ground motion using displacement maps only. Each input
sample consisted of a sequence of $L$ vertical displacement fields,
\begin{equation}
    \mathbf{X}^{\text{disp}}_n
    \in \mathbb{R}^{L \times 1 \times H \times W},
\end{equation}
and the network was tasked with predicting the next displacement map.
This single-modality setting provides a strong baseline and demonstrates
that deep spatio-temporal models can already capture long-term
subsidence trends and seasonal deformation patterns in EGMS data.

\subsubsection{Multimodal spatio-temporal representation}

To more explicitly expose physically meaningful factors that influence
ground deformation, we extend the input representation to three
complementary modalities, consistent with the feature construction
described in Section~\ref{sec:data}:

\begin{enumerate}
    \item \textbf{Dynamic displacement history.}
    The first modality is the normalised sequence of EGMS displacement
    maps $\mathbf{D}_{t-L+1:t}$, identical to the uni-modal setting.

    \item \textbf{Static deformation indicators.}
    From the \emph{training portion} of the EGMS time series at each
    grid cell we derive three summary statistics that characterise its
    long-term behaviour: (i) mean velocity, (ii) linear acceleration,
    and (iii) the amplitude of the dominant seasonal component. These
    indicators form a static feature tensor
    \begin{equation}
        \mathbf{S} \in \mathbb{R}^{3 \times H \times W},
    \end{equation}
    which is broadcast along the temporal dimension so that each time
    step receives the same spatially varying static context. Importantly,
    we recompute these indicators using only the training window for each
    tile, rather than using the EGMS L3 parameters fitted over the full
    2018--2022 series, to avoid any leakage of information from test
    epochs into the predictors.

    \item \textbf{Temporal encodings.}
    To inject information about acquisition date and seasonality directly
    into the model, we encode the day-of-year of each epoch $\tau$ using
    a harmonic representation,
    \begin{equation}
        \mathbf{h}_\tau =
        \bigl[\sin(2\pi \tau / T_{\text{year}}),\;
              \cos(2\pi \tau / T_{\text{year}})\bigr],
    \end{equation}
    where $T_{\text{year}} \approx 365.25$. The resulting two
    time features are tiled across the spatial grid, yielding a tensor
    of shape $2 \times H \times W$ per time step. This encoding allows
    the networks to learn seasonally dependent dynamics in a
    translation-invariant manner, as is common in time-series
    transformers and temporal convolutional models.
\end{enumerate}

Stacking all three modalities along the channel dimension leads to a
unified multimodal input of shape
\begin{equation}
\begin{aligned}
\mathbf{X}_n &\in \mathbb{R}^{L \times C \times H \times W},\\
C &= 1~(\text{displacement}) +
     3~(\text{static}) +
     2~(\text{time}) = 6.
\end{aligned}
\end{equation}

\subsubsection{Normalisation}

To stabilise optimisation and ensure comparability across space and
time, we apply modality-specific normalisation.

\paragraph{Displacement.}
For experiments where a model is trained and evaluated on a single tile,
we compute a per-pixel mean and standard deviation using only the
training portion of the time series,
\begin{equation}
\begin{aligned}
\mu_{d}(i,j) &=
    \frac{1}{T_{\text{train}}}
    \sum_{t=1}^{T_{\text{train}}} d_t(i,j),\\
\sigma_{d}(i,j) &=
    \sqrt{\frac{1}{T_{\text{train}}}
    \sum_{t=1}^{T_{\text{train}}}
    \bigl(d_t(i,j) - \mu_{d}(i,j)\bigr)^2 + \varepsilon}.
\end{aligned}
\end{equation}
and normalise as
$d_t^{\text{norm}}(i,j) = \bigl(d_t(i,j)-\mu_d(i,j)\bigr)/\sigma_d(i,j)$,
with $\varepsilon$ preventing division by zero. For cross-site
experiments, the statistics $\mu_{d}(i,j)$ and $\sigma_{d}(i,j)$ are
estimated on the source tile (E32N34) and applied unchanged to the input
displacement maps of the target tiles, so that the model sees test-time
input distributions scaled in the same way as during training.

\paragraph{Static indicators.}
For the static features we use channel-wise normalisation, subtracting
the spatial mean and dividing by the spatial standard deviation over all
pixels of each indicator. In all cases, these statistics are computed
from the static maps derived on the training window, and the resulting
affine transform is then applied to the static channels for the full
time series.

\paragraph{Temporal encodings.}
The sinusoidal encodings are naturally bounded in $[-1,1]$ and are used
without additional scaling.

At inference time, model predictions in normalised units are transformed
back to millimetres using the inverse of the per-pixel displacement
normalisation, enabling direct comparison with EGMS products and
geophysical thresholds.

\subsection{Model architectures}

We benchmark four families of spatio-temporal neural networks that are
representative of the main design paradigms used in geoscientific
forecasting: a convolutional recurrent baseline (CNN--LSTM), an
attention-augmented variant, a spatio-temporal graph convolutional
network (STGCN), and the proposed multimodal transformer.

\subsubsection{CNN--LSTM baseline}

The first baseline is a convolutional neural network followed by a long
short-term memory (CNN--LSTM) network, a widely used architecture for
modelling spatio-temporal sequences such as precipitation nowcasting and
traffic flow~\citep[e.g.][]{DBLP:journals/corr/ShiCWYWW15}. For each time step, a
stack of $2$-D convolutions with ReLU activations and spatial pooling
encodes the $C$-channel input map into a compact latent representation:
\begin{equation}
    \mathbf{z}_\tau = g_{\phi}\!\left(\mathbf{X}_n[\tau]\right)
    \in \mathbb{R}^{F},
    \quad \tau = t-L+1,\dots,t,
\end{equation}
where $g_{\phi}$ denotes the convolutional encoder and $F$ is the
flattened feature dimension. The resulting sequence
$(\mathbf{z}_{t-L+1},\dots,\mathbf{z}_t)$ is passed to a bidirectional
LSTM, which captures temporal dependencies and outputs a sequence of
hidden states $\mathbf{h}_\tau \in \mathbb{R}^{H_{\text{LSTM}}}$. We
aggregate temporal information by summing the hidden states and feed the
resulting context vector into a deconvolutional decoder composed of
fully connected and transposed convolution layers. This decoder
upsamples the latent representation back to the original spatial
resolution, producing the predicted map $\hat{d}_{t+1}(i,j)$.

\subsubsection{Attention-augmented CNN--LSTM}

To explicitly model the varying importance of different past epochs, we
construct an attention-augmented variant of the CNN--LSTM. After
obtaining the sequence of LSTM hidden states $\mathbf{h}_\tau$, we
compute a scalar attention score $e_\tau$ for each time step via a
shallow multilayer perceptron:
\begin{equation}
    e_\tau = \mathbf{w}^\top
    \tanh\!\left(\mathbf{W}_a \mathbf{h}_\tau + \mathbf{b}_a\right),
\end{equation}
normalise the scores with a softmax over time to obtain weights
$\alpha_\tau$, and form a context vector as the weighted sum
\begin{equation}
    \mathbf{c} = \sum_{\tau=t-L+1}^t \alpha_\tau \mathbf{h}_\tau.
\end{equation}
This context vector is then decoded to the output map using the same
deconvolutional decoder as in the baseline. The attention mechanism
allows the network to focus on epochs that are most informative for the
impending deformation, for instance strong seasonal peaks or recent
transients, which is particularly relevant for ground-motion analysis.

\subsubsection{Multimodal STGCN}

Graph-based methods offer an alternative view on spatio-temporal
dynamics by representing the study area as a graph and learning on its
nodes. We adopt a multimodal variant of the spatio-temporal graph
convolutional network (STGCN) originally proposed for traffic
forecasting~\citep{yu2017spatio}.

Each grid cell $(i,j)$ is treated as a node, and we construct an
undirected graph by connecting each node to its eight spatial neighbours
(queen adjacency), plus a self-loop. From this adjacency matrix
$\mathbf{A}$ we compute the symmetric normalised matrix
$\tilde{\mathbf{A}} = \mathbf{D}^{-1/2} (\mathbf{A} + \mathbf{I})
\mathbf{D}^{-1/2}$, where $\mathbf{D}$ is the degree matrix. The input
tensor is rearranged to shape $[B, C, N, L]$, where $N = H \times W$ is
the number of nodes and $B$ is the batch size.

An STGCN block consists of a temporal gated convolution, a graph
convolution, and a second temporal convolution. The temporal unit
applies $1$-D convolutions along the time dimension with a gated linear
unit (GLU) non-linearity:
\begin{equation}
    \mathbf{X}' = \sigma\!\left(\mathbf{W}_\text{in} * \mathbf{X}\right)
                  \odot
                  \left(\mathbf{W}_\text{out} * \mathbf{X}\right),
\end{equation}
where $*$ denotes convolution over time, $\sigma$ is the sigmoid
function, and $\odot$ is the element-wise product. Graph convolution is
then performed as
\begin{equation}
    \mathbf{X}'' = \mathbf{W}_g \mathbf{X}' \tilde{\mathbf{A}},
\end{equation}
where $\mathbf{W}_g$ is a learnable weight matrix. A residual connection
and layer normalisation stabilise training. Stacking two such blocks
yields a tensor of shape $[B, C', N, L]$, which is reshaped to
$[B, N, C' L]$ and passed through a fully connected layer to predict the
displacement at $t+1$ for each node. The node-wise predictions are
finally reshaped back to the image grid.

All modalities (displacement history, static indicators, temporal
encodings) are concatenated along the channel axis at the network input.
This enables the STGCN to exploit both local graph structure and
cross-modal interactions when propagating information over time. We
refer to this configuration as MM--STGCN in the experiments.

\subsubsection{Multimodal spatio-temporal transformer (ours)}

\begin{figure*}[t]
  \centering
  \includegraphics[width=\textwidth]{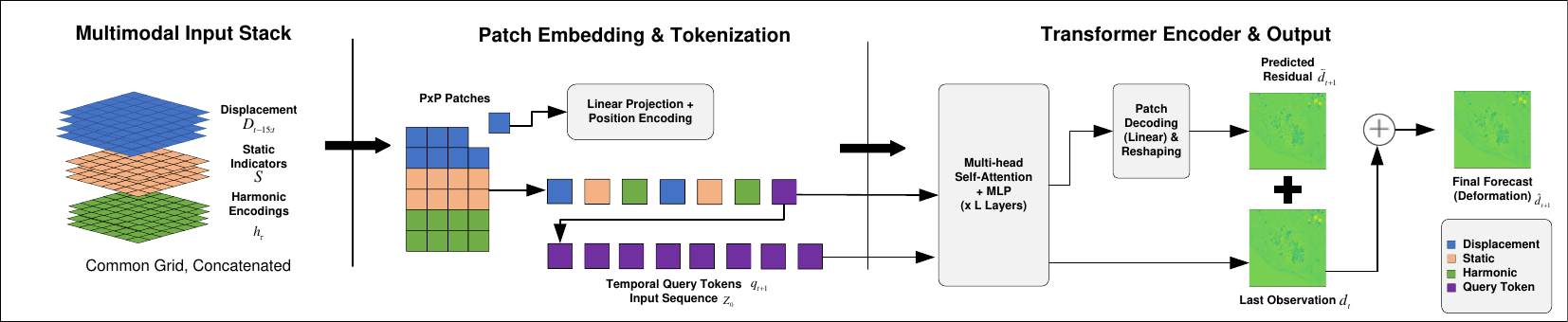}
  \caption{Conceptual architecture of the multimodal patch-based Transformer for EGMS ground-motion forecasting. For each tile and acquisition epoch, the vertical displacement history, static deformation indicators, and harmonic day-of-year encodings are stacked into a multi-channel grid. The grid is partitioned into fixed-size patches and linearly projected into patch embeddings, which are combined with spatial positional encodings. Together with temporal query tokens representing the forecast horizon ($t{+}1$), these embeddings are processed by a Transformer encoder with multi-head self-attention. The output query tokens are mapped back to patch-level residuals, assembled into a grid-based displacement map, and added to the last observed displacement. After de-normalisation, this yields the final single-step forecast in millimetres.}
  \label{fig:model_transformer}
\end{figure*}

Our main contribution is a multimodal spatio-temporal transformer
tailored to dense ground-deformation forecasting. The design is inspired
by the success of transformer architectures in sequence modelling and
vision tasks~\citep{DBLP:journals/corr/VaswaniSPUJGKP17}, but adapted to
handle EGMS displacement cubes in an efficient, patch-based manner. Figure~\ref{fig:model_transformer} shows the architecture of the Conceptual model.

\paragraph{Patch embedding and tokenisation.}
Each input tensor
$\mathbf{X}_n \in \mathbb{R}^{L \times C \times H \times W}$ is
partitioned into non-overlapping spatial patches of size $P \times P$.
For each time step and patch we flatten the $C \times P \times P$ values
into a vector and map it to a $D$-dimensional embedding using a linear
projection (with layer normalisation). This yields a sequence of
$L \times N_p$ tokens
$\mathbf{z}_{1:(L N_p)} \in \mathbb{R}^{D}$, where
$N_p = (H/P)\times(W/P)$ is the number of patches.

\paragraph{Temporal query tokens.}
Instead of autoregressively unrolling the transformer over time, we
append a set of learnable query tokens that correspond to the future
time steps we wish to predict (here $t+1$). The final input sequence to
the transformer encoder is
\begin{equation}
    \mathbf{Z}_0 =
    \bigl[\mathbf{z}_{1:(L N_p)},\; \mathbf{q}_{1:(L_\text{out}N_p)}\bigr]
    + \mathbf{P},
\end{equation}
where $\mathbf{q}$ denotes the query embeddings,
$L_\text{out}=1$ for single-step prediction, and $\mathbf{P}$ is a
learnable positional encoding that encodes both temporal index and patch
location.

\paragraph{Masked self-attention.}
We employ a standard multi-head self-attention encoder with $K$ layers.
In each layer, tokens are first normalised and then projected to
queries, keys and values, followed by scaled dot-product attention and a
feed-forward network:
\begin{align}
    \text{Attn}(\mathbf{Q},\mathbf{K},\mathbf{V})
    &= \text{softmax}\!\left(
        \frac{\mathbf{Q}\mathbf{K}^\top}{\sqrt{d_k}} + \mathbf{M}
       \right)\mathbf{V}, \\
    \mathbf{Z}_{\ell+1}
    &= \mathbf{Z}_{\ell}
       + \text{Attn}_\ell(\mathbf{Z}_\ell)
       + \text{FFN}_\ell(\mathbf{Z}_\ell),
\end{align}
where $\mathbf{M}$ is an additive attention mask. To preserve causality,
we mask attention from historical tokens to query tokens, while allowing
the queries to attend to all tokens in the sequence. This design lets
the future representations integrate information from all past patches
and modalities, while preventing leakage of predicted information back
into the encoder.

\paragraph{Patch decoding and residual prediction.}
After the final layer we extract the subset of tokens corresponding to
the future query positions and project each $D$-dimensional embedding
back to $C_\text{out} P^2$ pixels via a linear decoder and layer
normalisation. The resulting patch-wise predictions are rearranged to
form a displacement increment map $\tilde{d}_{t+1}(i,j)$. We adopt a
residual formulation in which the model predicts a correction to the
last observed displacement,
\begin{equation}
    \hat{d}_{t+1}(i,j) = d_t(i,j) + \tilde{d}_{t+1}(i,j),
\end{equation}
which empirically stabilises training and focuses the network on
modelling short-term increments rather than absolute displacements.

\subsection{Training objectives and optimisation}
\label{subsec:trainingoptim}

All models are trained to minimise a combination of pixel-wise
reconstruction losses and structural regularisers that encourage
physically plausible deformation patterns. Let $\hat{\mathbf{Y}}_n$ be
the predicted normalised displacement map and $\mathbf{Y}_n$ the
corresponding ground truth for sample $n$.

\paragraph{Reconstruction losses.}
Our primary objective is the mean absolute error (MAE) computed in
normalised units,
\begin{equation}
    \mathcal{L}_{\text{MAE}} =
    \frac{1}{B}\sum_{n=1}^{B}
    \bigl\| \hat{\mathbf{Y}}_n - \mathbf{Y}_n \bigr\|_1,
\end{equation}
where $B$ is the mini-batch size. To better reflect relative
discrepancies in millimetres, we also consider a relative error term
computed after de-normalisation:
\begin{equation}
    \mathcal{L}_{\text{rel}} =
    \frac{1}{B}\sum_{n=1}^{B}
    \frac{\bigl\| \hat{\mathbf{Y}}^{\text{mm}}_n
                 - \mathbf{Y}^{\text{mm}}_n \bigr\|_1}
         {\bigl\| \mathbf{Y}^{\text{mm}}_n \bigr\|_1 + \varepsilon},
\end{equation}
where $\mathbf{Y}^{\text{mm}}_n$ denotes the de-normalised map in
physical units.

\paragraph{Correlation and gradient regularisation.}
To promote coherence between predicted and observed spatial patterns, we
add two regularisers for the transformer model. First, we penalise
$1$ minus the Pearson correlation coefficient between the flattened
prediction and ground truth for each sample. Second, we introduce a
gradient loss that measures the MAE between Sobel-filtered horizontal
and vertical gradients of $\hat{\mathbf{Y}}^{\text{mm}}_n$ and
$\mathbf{Y}^{\text{mm}}_n$. This term encourages sharper subsidence
bowls and preserves local discontinuities, which are important for
interpreting deformation signals near infrastructure and faults.

\paragraph{Overall objective and optimisation.}
The overall loss is given by
\begin{equation}
    \mathcal{L} =
    \mathcal{L}_{\text{MAE}}
    + \lambda_{\text{rel}} \mathcal{L}_{\text{rel}}
    + \lambda_{\text{corr}} \mathcal{L}_{\text{corr}}
    + \lambda_{\text{grad}} \mathcal{L}_{\text{grad}},
\end{equation}
where the weighting factors $\lambda_{\cdot}$ are tuned on the
validation set. For the CNN--LSTM baselines and the STGCN we primarily
use $\mathcal{L}_{\text{MAE}}$
(i.e.\ $\lambda_{\text{rel}}=\lambda_{\text{corr}}=\lambda_{\text{grad}}=0$),
matching conventional practice in related work, while for the
transformer we exploit the full composite objective to maximise
structural fidelity.

All networks are trained with the AdamW optimiser, using a linearly
increasing learning-rate warm-up followed by cosine decay, gradient
clipping, and automatic mixed precision. We also maintain an exponential
moving average (EMA) of the model weights and use the EMA parameters for
validation and testing, which reduces the variance of the estimates and
leads to more stable convergence in practice. Early stopping based on
validation loss is employed to prevent overfitting, and model selection
for all architectures is carried out under the unified evaluation
protocol described in Section~\ref{sec:experiments}.

\subsection{Training regimes and cross-site setup}
\label{subsec:training-regimes}

The proposed preprocessing pipeline and multimodal representation
(Sections~\ref{sec:data}--\ref{sec:methods}) are applied consistently
to all EGMS tiles considered in this study. We distinguish three
training regimes, which correspond to the main groups of experiments
reported in Section~\ref{sec:experiments}.

\paragraph{In-tile training on eastern Ireland.}
For the core model comparison and ablation studies we train all
architectures (CNN--LSTM, CNN--LSTM+Attn, MM--STGCN and the proposed
multimodal transformer) on tile E32N34 covering eastern Ireland.
The data cube is split chronologically into 80\% training and 20\%
validation/test samples as described in
Sections~\ref{sec:data} and~\ref{sec:methods}. All models are trained
from scratch on this tile only, and hyperparameters are selected on
its validation set under the unified optimisation strategy of
Section~\ref{subsec:trainingoptim}.

We further perform an ablation in which all multimodal models are
trained under a simple SmoothL1 objective; see Section~\ref{subsec:loss_ablation}.

\paragraph{Cross-site generalisation across Europe.}
To assess whether the learned multimodal representation and transformer
architecture transfer across distinct deformation regimes, we perform
cross-site generalisation experiments. A single multimodal transformer
is trained on tile E32N34 (source tile) only, with early stopping based
on its validation loss. The resulting model is then applied without any
fine-tuning to a set of target tiles (E39N30, E44N23, E32N35, E48N24,
E58N17), which cover continuous subsidence, periodic deformation and
abrupt co-seismic displacement. For these experiments the
normalisation parameters for both displacement and static indicators are
estimated on the source tile and reused unchanged for all target tiles
(Section~\ref{subsec:input-repr}), so that the network sees input
statistics that are consistent with those encountered during training.

\paragraph{Site-specific training on representative deformation regimes.}
In addition to zero-shot cross-site transfer, we also investigate how
well the proposed transformer can be specialised to individual regions
when local training data are available. For each of the above tiles we
construct a local dataset using the same preprocessing, multimodal
feature construction and 80/20 temporal split as for E32N34.
A separate transformer is then trained from scratch on each tile and
evaluated on its held-out test portion. Comparing these site-specific
models with the cross-site generalisation results provides insight into
(i) how much performance is gained by local retraining, and
(ii) which deformation regimes are intrinsically more challenging to
predict given the EGMS time series.

\section{Experiments}
\label{sec:experiments}

\subsection{Experimental setup}

\subsubsection{Prediction task and data split}

All experiments address single–step prediction of ground deformation
using the EGMS Level~3 Upward component (Section~\ref{sec:data}). Given
a sequence of $T_{\mathrm{in}}$ past observations and auxiliary
variables, the models are trained to predict the displacement map at the
next epoch,
\begin{equation}
    \hat{\mathbf{D}}_{t+1} = f_\theta\!\left(\mathbf{X}_{t-T_{\mathrm{in}}+1:t}\right),
\end{equation}
where $\mathbf{X}_{t-T_{\mathrm{in}}+1:t}$ denotes either the
displacement–only or the multimodal input described in
Section~\ref{sec:methods}, and $\theta$ are learnable parameters.

Training samples are constructed by sliding a fixed–length window along
each EGMS time series and discarding incomplete windows at the beginning
and end of the acquisition period. For every tile, the time series are
split chronologically: the first $80\,\%$ of epochs are used for
training and the remaining $20\,\%$ for validation and testing. All
reported numbers are computed on this held–out temporal segment, so that
models are always evaluated on future epochs that were never seen during
training.

For the core model comparison we consider tile E32N34 (eastern Ireland).
Cross–site experiments follow the regimes defined in
Section~\ref{subsec:training-regimes}, covering additional EGMS tiles
across Europe that represent distinct deformation types
(continuous subsidence, seasonal motion and abrupt co–seismic offsets).

\subsubsection{Input configurations}

We evaluate two input configurations:

\begin{enumerate}
    \item \textbf{Displacement–only (unimodal).}
    The input at each time step consists solely of the EGMS displacement
    map, normalised per–pixel as described in
    Section~\ref{subsec:input-repr}. This setting mirrors the
    configuration investigated in our previous work
    \citep{yao2025deeplearningapproachspatiotemporal} and serves as a
    reference for pure time–series deformation forecasting.

    \item \textbf{Multimodal (proposed).}
    In addition to the displacement map, the input stack contains three
    static fields (mean velocity, linear acceleration and seasonal
    amplitude) and a two–dimensional temporal encoding
    $(\sin\phi_t,\cos\phi_t)$ of the day–of–year. All channels are
    jointly interpolated to a common $64\times64$ grid and normalised.
    The resulting six–channel tensor allows the models to exploit
    long–term kinematics and seasonal cycles even though the prediction
    horizon is one step.
\end{enumerate}

Unless otherwise stated, all architectures use the same preprocessing
pipeline and the same temporal train/validation split in both
configurations. The only difference between the two settings is the
number of channels provided to the networks.

\subsubsection{Training and evaluation protocol}

All models are implemented in PyTorch. For the CNN--LSTM and
CNN--LSTM+Attention baselines we follow common settings for
spatio–temporal sequence modelling \citep{DBLP:journals/corr/ShiCWYWW15}.
The MM--STGCN uses two spatio–temporal blocks with an 8–neighbour
regular grid graph over the $64\times64$ pixels, as detailed in
Section~\ref{sec:methods}. The proposed multimodal transformer is
trained with the composite loss of
Section~\ref{subsec:trainingoptim}, combining per–pixel $\ell_1$ loss,
relative error in millimetres, correlation and gradient regularisation in
a residual–prediction setting.

All networks are optimised with Adam or AdamW, using learning–rate
warm–up followed by cosine decay, gradient clipping and automatic mixed
precision. We maintain an exponential moving average (EMA) of the model
parameters and use the EMA weights for validation and testing. Early
stopping on the validation loss is employed in all cases.

We report root–mean–square error (RMSE), mean absolute error (MAE) and
coefficient of determination ($R^2$) computed over all pixels and
time steps in the test set. In addition, we use threshold–based
accuracies:
\begin{itemize}
    \item relative thresholds $\mathrm{Acc@10\%,20\%,50\%}$, defined as
    the fraction of pixels whose absolute error is below
    $10\,\%,20\,\%,50\,\%$ of the true displacement magnitude;
    \item absolute thresholds in millimetres,
    $\mathrm{Acc@1mm,0.5mm,0.2mm,0.1mm}$, defined as the fraction of
    pixels whose absolute error is below the corresponding level.
\end{itemize}
For qualitative assessment we also compute the structural similarity
index (SSIM) and Pearson correlation between predicted and reference
maps, and visualise scatter plots and residual diagnostics.

\subsection{Model comparison on eastern Ireland (E32N34)}
\label{sec:results-e32n34}

\subsubsection{Displacement–only baseline}

We first compare two simple geodetic baselines with the four deep
architectures when only displacement maps are used as input. The
classical baselines operate independently on each pixel time series:
(i) a linear trend model fitted over the training period and
extrapolated one step ahead, and (ii) a linear trend plus seasonal
harmonic model using a single annual sinusoid. Table~\ref{tab:results_unimodal}
summarises the single--step prediction performance on tile E32N34 in
this unimodal setting.

\begin{table}[t]
    \small
    \centering
    \setlength{\tabcolsep}{1mm}
    \caption{Single--step prediction performance on tile E32N34 with
    displacement--only inputs. Best values are highlighted in bold.}
    \label{tab:results_unimodal}
    \begin{tabular}{lccc}
        \toprule
        Model & RMSE [mm] & MAE [mm] & $R^2$ \\
        \midrule
        Linear trend                  & 2.1041 & 1.1933 & 0.8428 \\
        Linear + seasonal harmonic    & 2.0168 & 1.1451 & 0.8555 \\
        CNN--LSTM                     & 2.0239 & 1.3442 & 0.8621 \\
        CNN--LSTM+Attn                & 1.8230 & 1.1268 & 0.8881 \\
        STGCN                         & \textbf{1.0410} & \textbf{0.6957} & \textbf{0.9635} \\
        Transformer (ours)            & 1.5873 & 0.9389 & 0.9152 \\
        \bottomrule
    \end{tabular}
\end{table}

The two classical baselines provide a useful lower bound on performance:
the linear trend and linear--plus--seasonal models reach
$\mathrm{RMSE}\approx 2.1$~mm and $\mathrm{RMSE}\approx 2.0$~mm,
respectively, with $R^2$ between $0.84$ and $0.86$. All deep models
improve upon these geodetic baselines in terms of global error and
explained variance. Among the neural architectures, STGCN clearly
provides the strongest performance when only displacement history is
available, achieving an RMSE of $1.04$~mm and an $R^2$ of $0.96$. The
proposed transformer attains an RMSE of $1.59$~mm and $R^2=0.92$,
outperforming both CNN--LSTM variants but remaining behind STGCN in this
configuration.

Threshold accuracies in Table~\ref{tab:thr_unimodal} provide additional
insight. The classical baselines already achieve reasonable fractions of
pixels within 1~mm absolute error (around $60$--$61\%$), but still lag
behind STGCN and the transformer at the strictest thresholds. STGCN
attains the highest fraction of pixels within $1$~mm and $0.5$~mm, while
the transformer yields the best performance in the most demanding
$\mathrm{Acc@0.1mm}$ bin.

\begin{table*}[t]
    \centering
    \caption{Threshold accuracies on tile E32N34 for displacement--only
    inputs. Values are given in percent.}
    \label{tab:thr_unimodal}
    \begin{tabular}{lccccccc}
        \toprule
        Model & Acc@10\% & Acc@20\% & Acc@50\% & Acc@1mm & Acc@0.5mm & Acc@0.2mm & Acc@0.1mm \\
        \midrule
        Linear trend               & 33.77 & 46.29 & 69.38 & 59.99 & 41.65 & 28.43 & 23.75 \\
        Linear + seasonal harmonic & \textbf{34.72} & \textbf{47.65} & \textbf{70.26} & 61.48 & 42.81 & 28.89 & 24.01 \\
        CNN--LSTM                  & 11.94 & 23.37 & 48.74 & 54.37 & 32.24 & 13.76 &  6.90 \\
        CNN--LSTM+Attn             & 14.19 & 27.50 & 53.57 & 61.26 & 42.70 & 26.45 & 15.69 \\
        STGCN                      & 24.66 & 40.83 & 61.52 & \textbf{76.15} & \textbf{54.51} & \textbf{34.32} & 10.78 \\
        Transformer (ours)         & 16.16 & 31.92 & 59.24 & 67.58 & 46.76 & 27.97 & \textbf{16.87} \\
        \bottomrule
    \end{tabular}
\end{table*}

Figure~\ref{fig:map_unimodal} shows a representative example of
predicted displacement maps for all models. All architectures reproduce
the broad coastal subsidence bowl, but STGCN and the transformer exhibit
higher structural fidelity than the CNN--LSTM baselines, with SSIM and
map--wise correlation close to $0.98$ on this epoch and clearly sharper
structures than those implied by the pixel--wise classical regressions.

\subsubsection{Multimodal conditioning on static and temporal covariates}

We next enable the full multimodal input (displacement + static
kinematics + temporal encoding) for all models. The corresponding
quantitative results on E32N34 are reported in
Table~\ref{tab:results_multimodal}.

\begin{table}[t]
    \centering
    \small
    \setlength{\tabcolsep}{1mm} 
    \caption{Single–step prediction performance on tile E32N34 with
    multimodal inputs (displacement + static kinematics + temporal
    encoding).}
    \label{tab:results_multimodal}
    \begin{tabular}{lccc}
        \toprule
        Model & RMSE [mm] & MAE [mm] & $R^2$ \\
        \midrule
        CNN--LSTM (Multi)         & 1.9704 & 1.1721 & 0.8693 \\
        CNN--LSTM+Attn (Multi)    & 1.9112 & 1.1579 & 0.8770 \\
        MM--STGCN                 & 1.4465 & 0.7299 & 0.9295 \\
        Transformer (ours, Multi) & \textbf{0.9007} & \textbf{0.5738} & \textbf{0.9727} \\
        \bottomrule
    \end{tabular}
\end{table}

Augmenting the input with static and temporal covariates substantially
changes the relative ranking. The multimodal transformer now achieves
the best performance, with an RMSE of $0.90$~mm and $R^2=0.97$,
outperforming the multimodal STGCN (RMSE $1.45$~mm, $R^2=0.93$) and both
CNN--LSTM variants. Compared with its displacement–only counterpart, the
transformer’s RMSE decreases by roughly $40\,\%$ and the explained
variance increases from $0.92$ to $0.97$.

Table~\ref{tab:thr_multimodal} reports the threshold accuracies for the
multimodal setting. All architectures benefit from the enriched inputs,
but the gains are most pronounced for the transformer: Acc@10\,\%
almost triples (from $16.2\,\%$ to $47.2\,\%$), and the proportion of
predictions within $1$~mm increases from $67.6\,\%$ to $81.0\,\%$.

\begin{table*}[t]
    \centering
    \caption{Threshold accuracies on tile E32N34 for multimodal inputs.
    Values are given in percent.}
    \label{tab:thr_multimodal}
    \begin{tabular}{lccccccc}
        \toprule
        Model & Acc@10\% & Acc@20\% & Acc@50\% & Acc@1mm & Acc@0.5mm & Acc@0.2mm & Acc@0.1mm \\
        \midrule
        CNN--LSTM (Multi)         & 25.74 & 43.51 & 71.44 & 59.98 & 41.83 & 28.50 & 23.84 \\
        CNN--LSTM+Attn (Multi)    & 29.96 & 44.43 & 72.03 & 59.84 & 41.59 & 28.35 & 23.67 \\
        MM--STGCN                 & 28.89 & 58.21 & 80.70 & 75.05 & 53.29 & 33.78 & 26.46 \\
        Transformer (ours, Multi) & \textbf{47.22} & \textbf{63.69} & \textbf{82.37} & \textbf{81.01} & \textbf{59.31} & \textbf{37.19} & \textbf{28.39} \\
        \bottomrule
    \end{tabular}
\end{table*}

Qualitative maps in Figure~\ref{fig:map_multimodal} illustrate the
impact of multimodal conditioning. The transformer predictions are
visually almost indistinguishable from the EGMS reference, with SSIM and
map–wise correlation close to $0.99$ and the smallest residuals in both
the coastal subsidence bowl and the more stable inland areas. The
multimodal STGCN also yields sharper structures than its unimodal
version, but exhibits a slight amplitude bias in the highest–subsidence
zones. Residual diagnostics and binned error statistics (see
Figure~\ref{fig:analysis_multimodal}) show that the multimodal
transformer maintains stable accuracy across the full deformation range,
with narrow residual distributions even in the largest–magnitude bins.

\subsection{Cross–site experiments across Europe}
\label{sec:cross-site-results}

We now turn to the six EGMS tiles introduced in
Section~\ref{sec:data}, which were selected to span three deformation
regimes: continuous subsidence (E39N30, E44N23), seasonal deformation
(E32N34, E32N35) and abrupt co–seismic offsets (E48N24, E58N17). All
experiments in this subsection use the multimodal input configuration
and the transformer architecture.

\subsubsection{Site–specific multimodal training}

In the first set of experiments, we train a separate multimodal
transformer from scratch on each tile, following the site–specific
regime in Section~\ref{subsec:training-regimes}. Table~\ref{tab:sitespecific}
summarises the single–step performance on the held–out test portion of
each time series.

\begin{table*}[t]
    \centering
    \caption{Site–specific multimodal transformer performance for
    representative EGMS tiles. Each model is trained and evaluated on a
    single tile using the 80/20 temporal split.}
    \label{tab:sitespecific}
    \begin{tabular}{lccccc}
        \toprule
        Tile ID & Deformation type & RMSE [mm] & MAE [mm] & $R^2$ & Acc@1mm [\%] \\
        \midrule
        E32N34 & Seasonal / mixed          & 0.9007 & 0.5738 & 0.9727 & 81.01 \\
        E32N35 & Seasonal                  & 1.0352 & 0.6290 & 0.9251 & 79.22 \\
        E39N30 & Continuous subsidence     & 1.1199 & 0.7580 & 0.9516 & 75.56 \\
        E44N23 & Continuous subsidence     & 1.4100 & 0.9089 & 0.9508 & 69.78 \\
        E58N17 & Co–seismic (Croatia)      & 4.0497 & 2.5462 & 0.9882 & 48.92 \\
        \bottomrule
    \end{tabular}
\end{table*}

Across all tiles and deformation regimes, the site–specific multimodal
transformer attains high coefficients of determination
($R^2 \ge 0.93$ for all seasonal and continuous–subsidence sites,
$R^2 \approx 0.99$ for the co–seismic case E58N17), with between
$70\,\%$ and $80\,\%$ of pixels within $1$~mm error for the seasonal and
continuous sites. In co–seismic areas, absolute errors in millimetres
are higher due to the larger signal amplitudes, but the relative errors
remain small and $R^2$ values very close to~1, indicating that the
transformer can reproduce both the location and the shape of abrupt
displacement fields.

\subsubsection{Cross–site generalisation from eastern Ireland}

To quantify cross–site generalisation, we next train a single multimodal
transformer only on tile E32N34 (eastern Ireland) and evaluate it
without any fine–tuning on the remaining five tiles, as described in
Section~\ref{subsec:training-regimes}. For these experiments, the
normalisation parameters (per–pixel displacement statistics and
per–channel static descriptors) are estimated on E32N34 and reused for
all other tiles, so that the model sees input statistics consistent with
its training environment.

Table~\ref{tab:crosssite} reports the zero–shot performance of this
E32N34–trained transformer across all six tiles.

\begin{table*}[t]
    \centering
    \caption{Cross--site generalisation of a multimodal Transformer and a multimodal STGCN trained only on tile E32N34 and evaluated without fine--tuning on the other five tiles. All inputs are normalised using statistics from E32N34.}
    \label{tab:crosssite}
    \begin{tabular}{l l l c c c c}
        \toprule
        Tile ID & Deformation type & Model
                & RMSE [mm] & MAE [mm] & $R^2$ & Acc@1mm [\%] \\
        \midrule
        \multirow{2}{*}{E32N34 (source)}
          & \multirow{2}{*}{Seasonal / mixed}
          & Transformer (ours)   & 0.7353 & 0.4092 & 0.9811 & 87.37 \\
          &                      & MM--STGCN           & 1.0178 & 0.6684 & 0.9651 & 76.77 \\
        \midrule
        \multirow{2}{*}{E32N35}
          & \multirow{2}{*}{Seasonal}
          & Transformer (ours)   & 0.9996 & 0.5743 & 0.9301 & 80.40 \\
          &                      & MM--STGCN           & 1.5341 & 1.0772 & 0.8404 & 58.82 \\
        \midrule
        \multirow{2}{*}{E39N30}
          & \multirow{2}{*}{Continuous subside nce}
          & Transformer (ours)   & 0.9699 & 0.5799 & 0.9637 & 81.20 \\
          &                      & MM--STGCN           & 1.8133 & 1.3437 & 0.8778 & 48.54 \\
        \midrule
        \multirow{2}{*}{E44N23}
          & \multirow{2}{*}{Continuous subsidence}
          & Transformer (ours)   & 1.2670 & 0.7826 & 0.9603 & 74.93 \\
          &                      & MM--STGCN           & 2.1748 & 1.6278 & 0.8872 & 40.73 \\
        \midrule
        \multirow{2}{*}{E48N24}
          & \multirow{2}{*}{Co--seismic (Samos)}
          & Transformer (ours)   & 3.1777 & 1.0219 & 0.9729 & 73.19 \\
          &                      & MM--STGCN           & 6.6834 & 2.9453 & 0.8807 & 32.23 \\
        \midrule
        \multirow{2}{*}{E58N17}
          & \multirow{2}{*}{Co--seismic (Croatia)}
          & Transformer (ours)   & 1.5520 & 0.8047 & 0.9983 & 71.40 \\
          &                      & MM--STGCN           & 8.3652 & 6.3216 & 0.9513 & 15.08 \\
        \bottomrule
    \end{tabular}
\end{table*}

Several observations emerge. First, the cross–site transformer achieves
$R^2 \ge 0.93$ on all target tiles, despite never seeing their data
during training and using normalisation statistics from a different
region. For the seasonal and continuous–subsidence sites, RMSE values
remain close to or below $1$~mm and Acc@1mm ranges between $74\,\%$ and
$81\,\%$, only moderately lower than the site–specific models in
Table~\ref{tab:sitespecific}. Second, the transformer also attains high overall $R^2$ and competitive
threshold accuracies on the two co--seismic tiles (E48N24 and E58N17).
However, these aggregate scores must be interpreted with care.
Our task is one-step-ahead nowcasting: for each prediction
$\hat{d}_{t+1\,|\,1:t}$ the input history window contains observations only up to epoch $t$.
Therefore, the abrupt co--seismic offset at the event epoch cannot be
anticipated unless it has already been observed in the input sequence.
In practice, the event enters the input window only \emph{after} the first
post-event acquisition, and the model performance over the long post-seismic
period largely reflects its ability to remain numerically stable and
to update its forecasts once the discontinuity has occurred.

\begin{figure*}[htbp]
  \centering
  \subfigure[]{
    \includegraphics[width=0.48\textwidth]{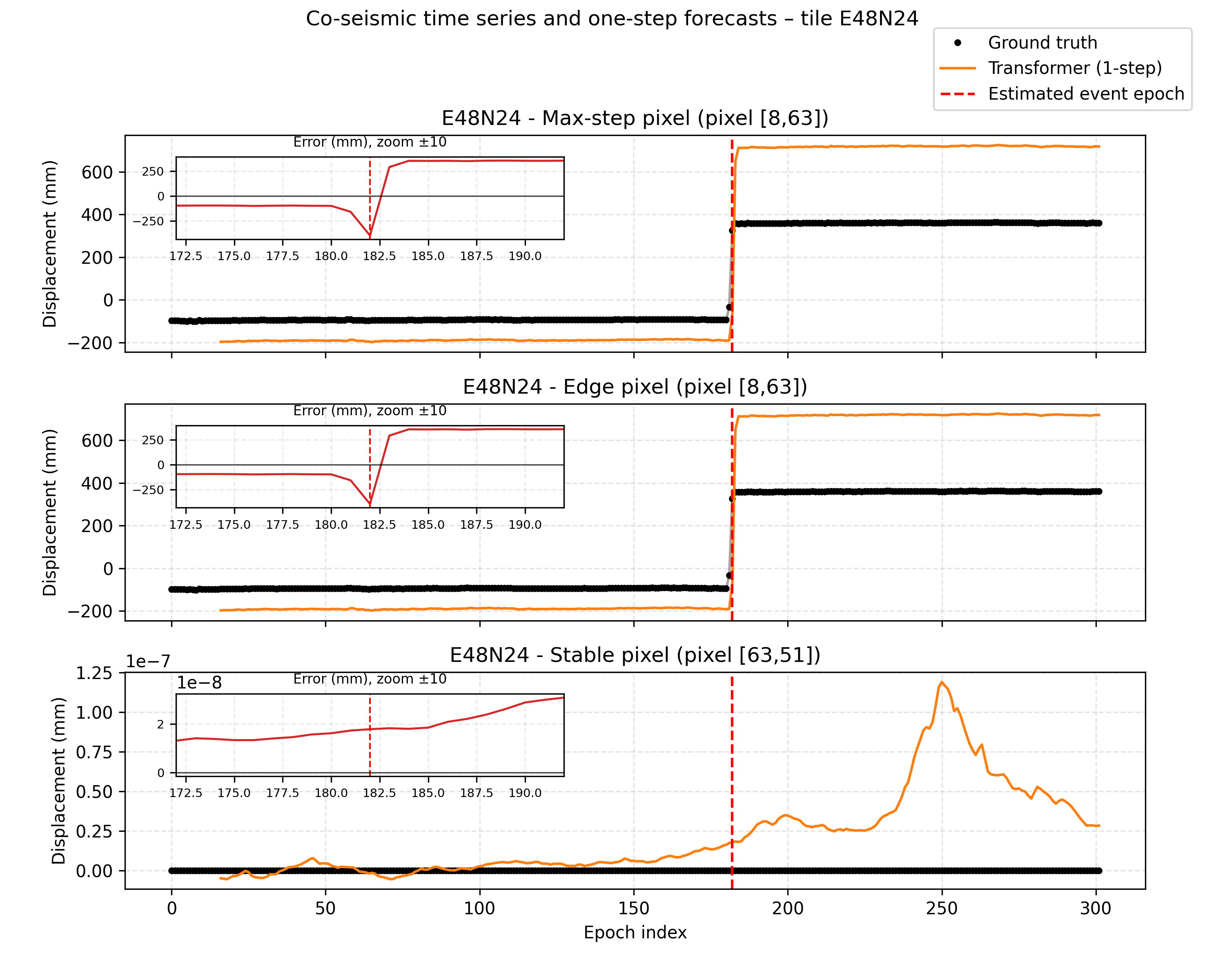}
    \label{fig:analysis_E48N24}
  }
  \hfill
  \subfigure[]{
    \includegraphics[width=0.48\textwidth]{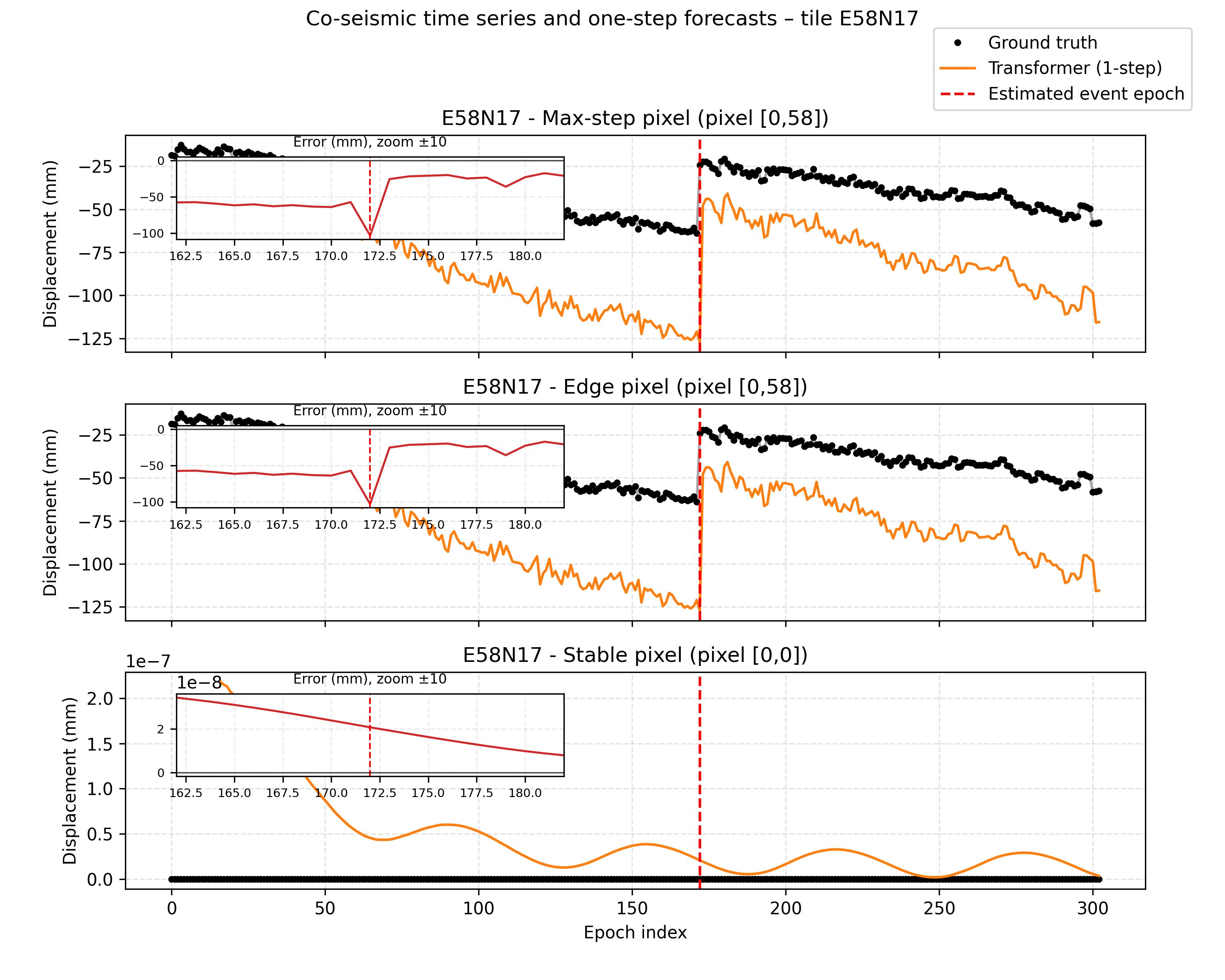}
    \label{fig:analysis_E58N17}
  }
  \caption{Event-centred interpretation of co--seismic ``forecasting'' in the zero-shot setting.
  For each co--seismic tile (E48N24: Samos; E58N17: Croatia), we show three representative pixels
  (largest step, step-edge, and a stable/background pixel). Black markers denote EGMS displacement
  time series, and the orange curve is the multimodal Transformer one-step-ahead nowcast
  $\hat{d}_{t+1\,|\,1:t}$ produced by sliding the input history window through time.
  The red dashed line indicates the estimated co--seismic epoch (largest absolute one-step change),
  and the inset in each panel plots the prediction error (prediction minus truth) in a $\pm 10$
  epoch window around the event. As expected for a single-step forecaster trained on smooth
  deformation histories, the model does not anticipate the abrupt jump and exhibits a sharp error
  peak at the co--seismic epoch; after the discontinuity is observed and enters the conditioning
  history, the predictions remain stable and track the post-event state.}
  \label{fig:coseismic_timeseries}
\end{figure*}

To make this explicit, Figure~\ref{fig:coseismic_timeseries} reports
representative pixel-wise time series for both co--seismic tiles, together with
their one-step predictions and the event-centred error curves.
The plots show an unavoidable error peak around the estimated event epoch,
followed by stable behaviour afterwards once the post-seismic level becomes
part of the conditioning history. Hence, the transfer results on co--seismic
tiles should be understood as \emph{robustness to strong discontinuities and
rapid post-event updating}, rather than the prediction of earthquake occurrence.

\begin{figure*}[t]
  \centering
  \subfigure[E32N34 (seasonal / mixed), trained on E32N34]{
    \includegraphics[width=0.3\textwidth]{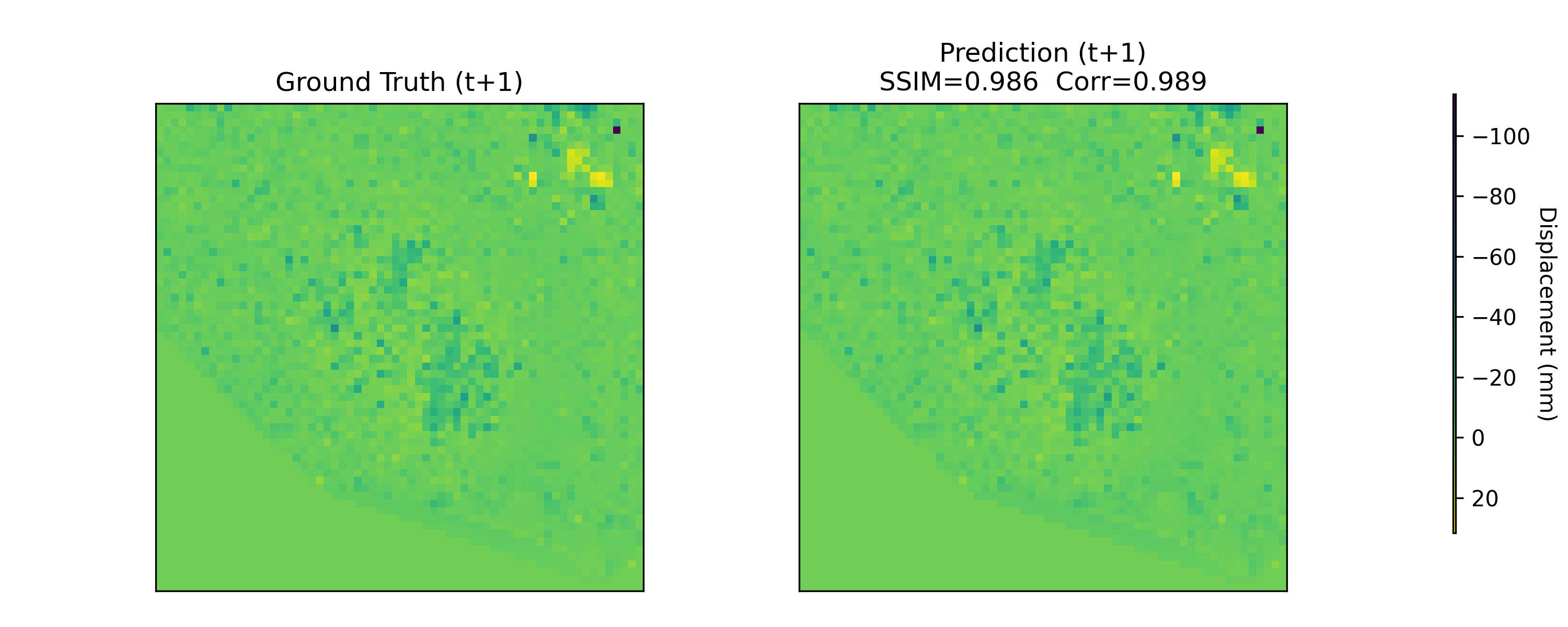}
    \label{fig:gen_E32N34}
  }
  \hfill
  \subfigure[E32N35 (seasonal), zero-shot from E32N34]{
    \includegraphics[width=0.3\textwidth]{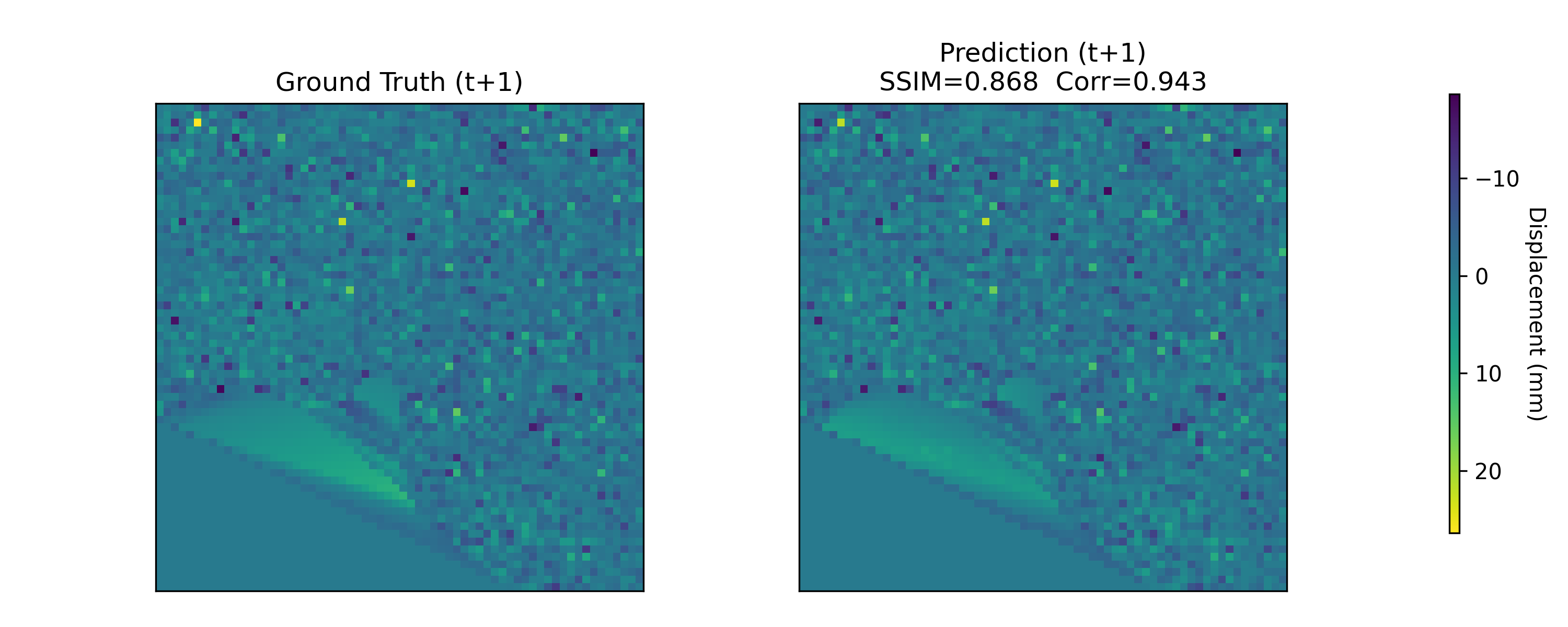}
    \label{fig:gen_E32N35}
  }
  \hfill
  \subfigure[E39N30 (continuous subsidence), zero-shot from E32N34]{
    \includegraphics[width=0.3\textwidth]{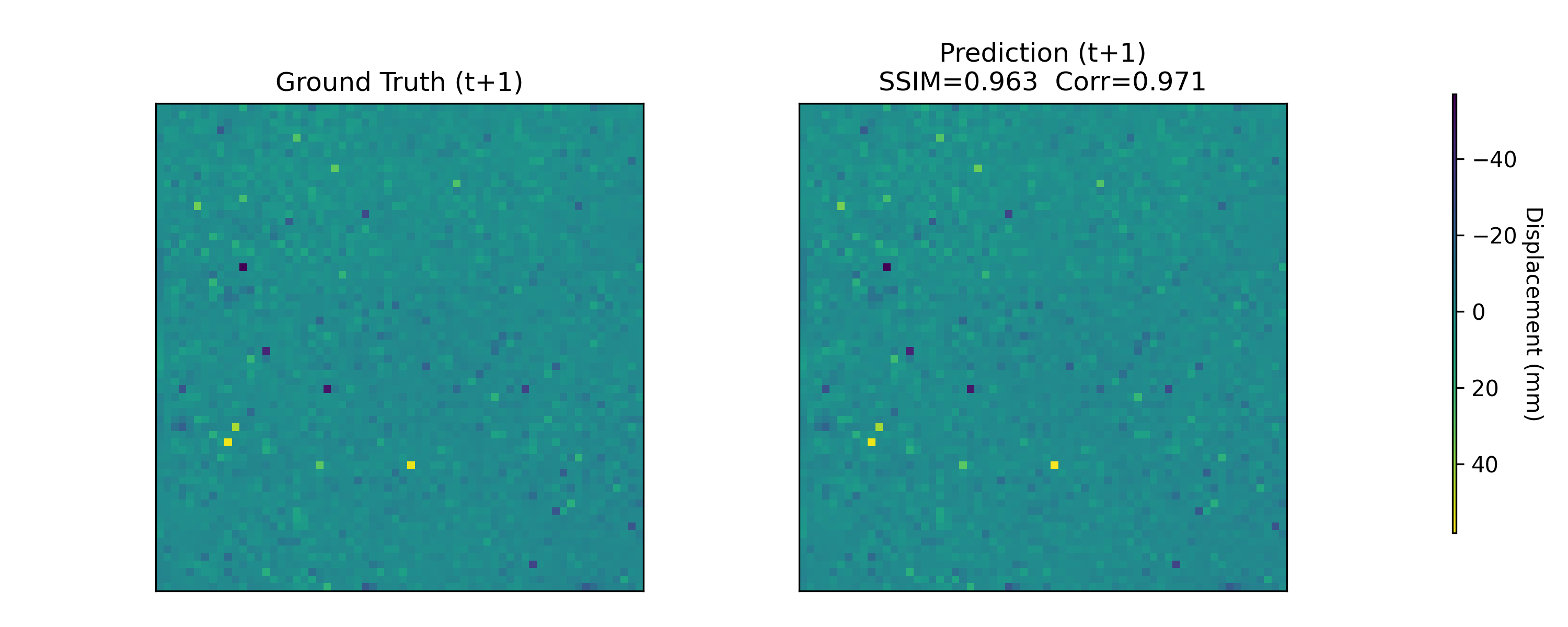}
    \label{fig:gen_E39N30}
  }

  \vspace{0.5em}

  \subfigure[E44N23 (continuous subsidence), zero-shot from E32N34]{
    \includegraphics[width=0.3\textwidth]{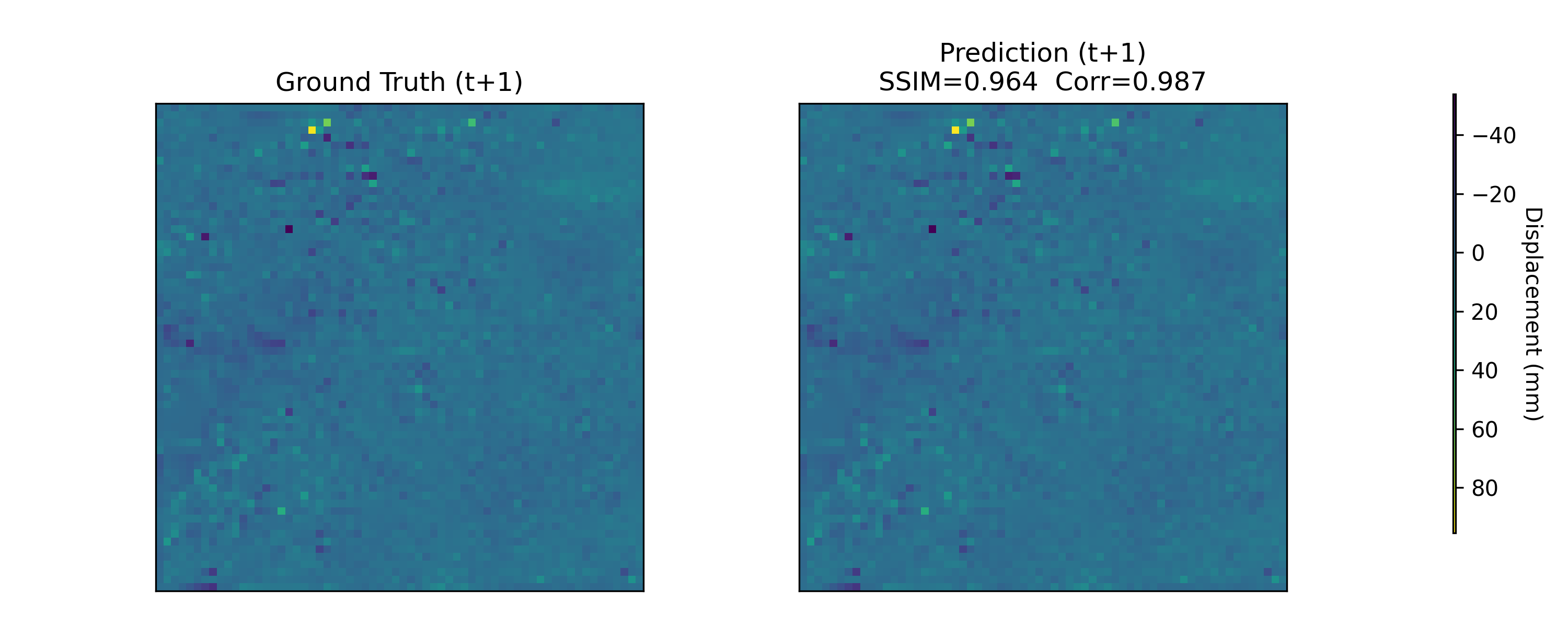}
    \label{fig:gen_E44N23}
  }
  \hfill
  \subfigure[E48N24 (co-seismic, Samos), zero-shot from E32N34]{
    \includegraphics[width=0.3\textwidth]{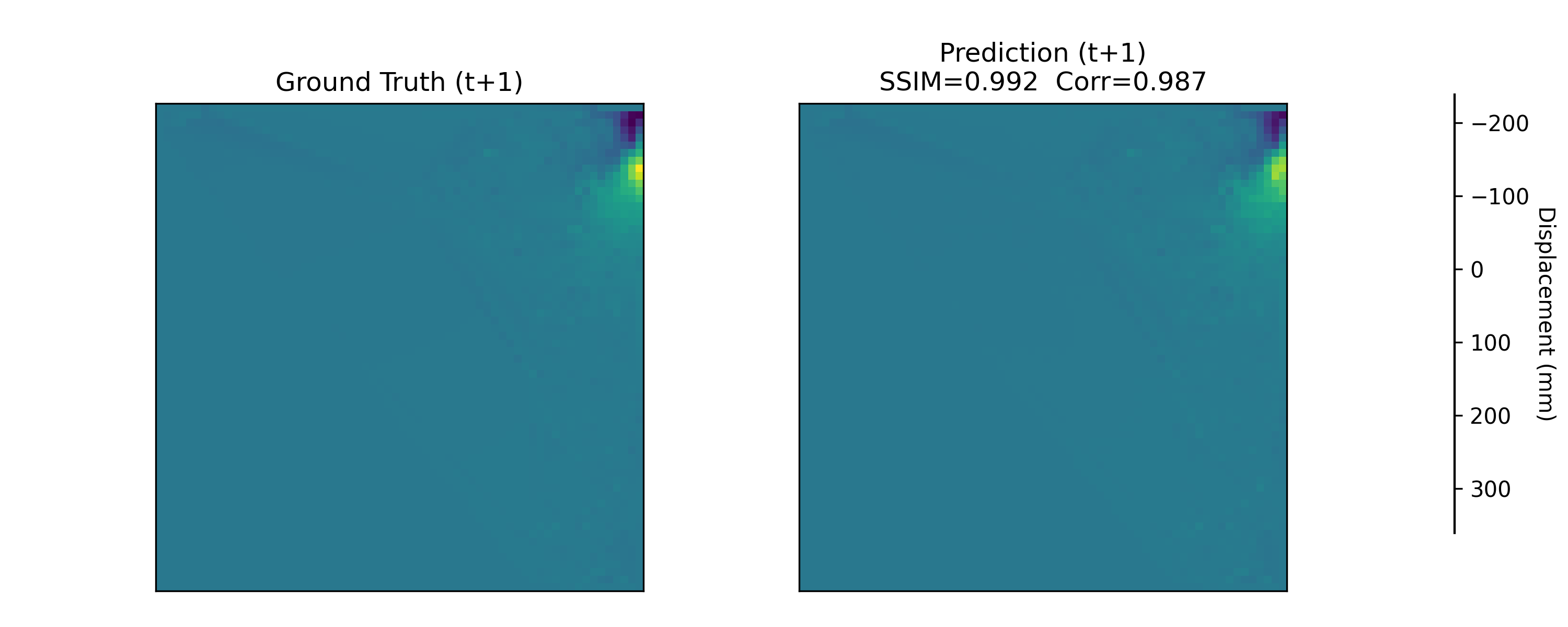}
    \label{fig:gen_E48N24}
  }
  \hfill
  \subfigure[E58N17 (co-seismic, Croatia), zero-shot from E32N34]{
    \includegraphics[width=0.3\textwidth]{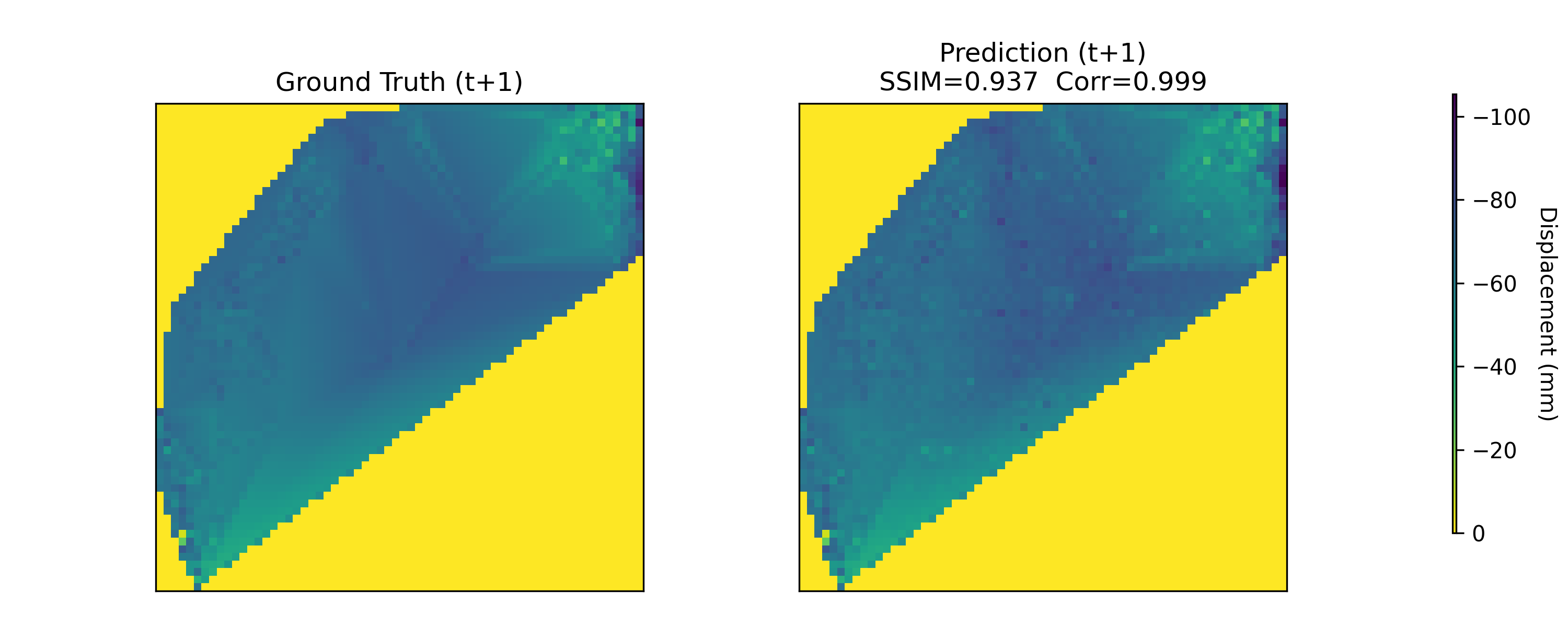}
    \label{fig:gen_E58N17}
  }

\caption{Qualitative examples of \emph{zero-shot} cross-tile single-step ($t{+}1$) forecasts.
The multimodal Transformer is trained only on tile E32N34 (eastern Ireland) and applied to the other tiles
\emph{without fine-tuning}. Each panel shows the ground truth at $t{+}1$ (left) and the prediction (right),
covering seasonal deformation (E32N35), continuous subsidence (E39N30, E44N23) and co-seismic offsets (E48N24, E58N17).}

  \label{fig:maps_generalisation}
\end{figure*}

Figure~\ref{fig:maps_generalisation} illustrates qualitative examples of
\emph{zero-shot cross-tile transfer} for single-step ($t{+}1$) forecasting.
A multimodal Transformer trained only on tile E32N34 is applied to the other
tiles \emph{without any fine-tuning or retraining}, so the figure visualises
transferability rather than site-specific accuracy.

To verify that the strong cross--site performance is not a generic
property of the data or of any reasonably regularised model, we also
evaluate a multimodal STGCN trained only on tile E32N34 and applied
zero--shot to the same set of target tiles (Table~\ref{tab:crosssite}).
In this setting the STGCN exhibits noticeably weaker generalisation:
for the continuous--subsidence tiles E39N30 and E44N23, RMSE increases
to 1.8 to 2.2~mm and Acc@1mm drops to 41--49\%, while for the co--seismic
tiles the errors become substantially larger (RMSE $>6$~mm on E48N24 and
$>8$~mm on E58N17) and fewer than one third of pixels fall within 1~mm
of the reference displacement.

Compared to the transformer in Table~\ref{tab:crosssite}, which maintains
RMSE close to or below 1~mm and Acc@1mm between 71\% and 81\% on most
tiles, these results indicate that cross--site generalisation is not
trivial and strongly depends on the model architecture. The multimodal
transformer retains much of its site--specific skill when transferred
to unseen regions, whereas the STGCN, despite being competitive in the
single--site experiments, degrades more severely when exposed to
distribution shifts in deformation regime and noise characteristics.
\subsection{Summary of experimental findings}

Overall, the experiments can be summarised as follows:

\begin{itemize}
    \item On a single tile (E32N34), STGCN is the strongest architecture
    when only displacement maps are available, while the proposed
    multimodal transformer becomes the top performer once static and
    temporal covariates are included, reducing RMSE by about $40\,\%$
    relative to its unimodal counterpart and surpassing STGCN in both
    global metrics and threshold accuracies.

    \item Across six EGMS tiles spanning continuous, seasonal and
    co--seismic deformation, site–specific multimodal transformers
    systematically achieve high $R^2$ and robust Acc@1mm, confirming
    that the architecture can adapt to very different deformation
    regimes.

    \item A single multimodal transformer trained on eastern Ireland
    generalises surprisingly well to all other tiles without
    fine–tuning, maintaining $R^2 \ge 0.93$ and Acc@1mm in the
    $70\text{--}80\,\%$ range. Performance for co–seismic tiles is
    particularly strong in relative terms.

\end{itemize}

These empirical results form the basis for the broader interpretation
and implications discussed in Section~\ref{sec:discussion}.


\begin{figure*}[t]
  \centering
  \includegraphics[width=\textwidth]{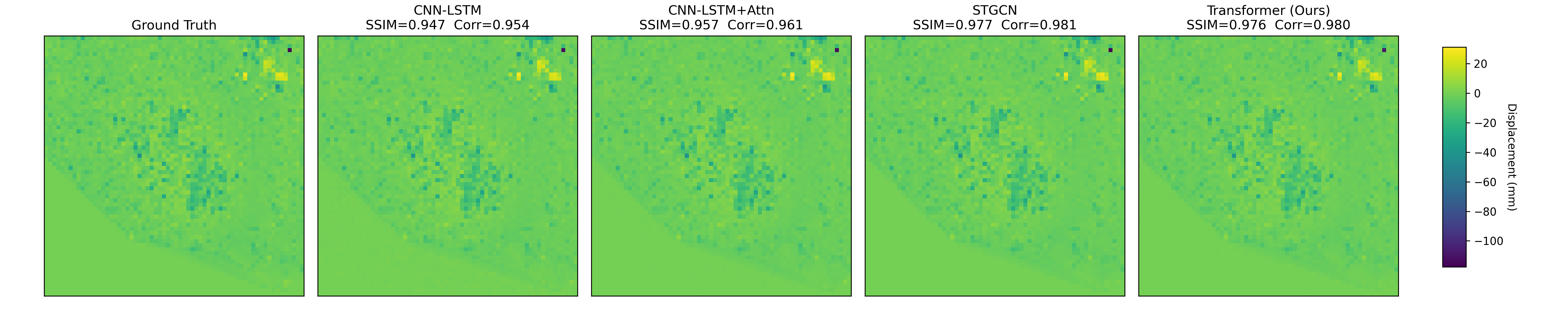}
  \caption{Single–step predicted displacement map comparison for the
  displacement–only setting on tile E32N34.}
  \label{fig:map_unimodal}
\end{figure*}

\begin{figure*}[t]
  \centering
  \includegraphics[width=\textwidth]{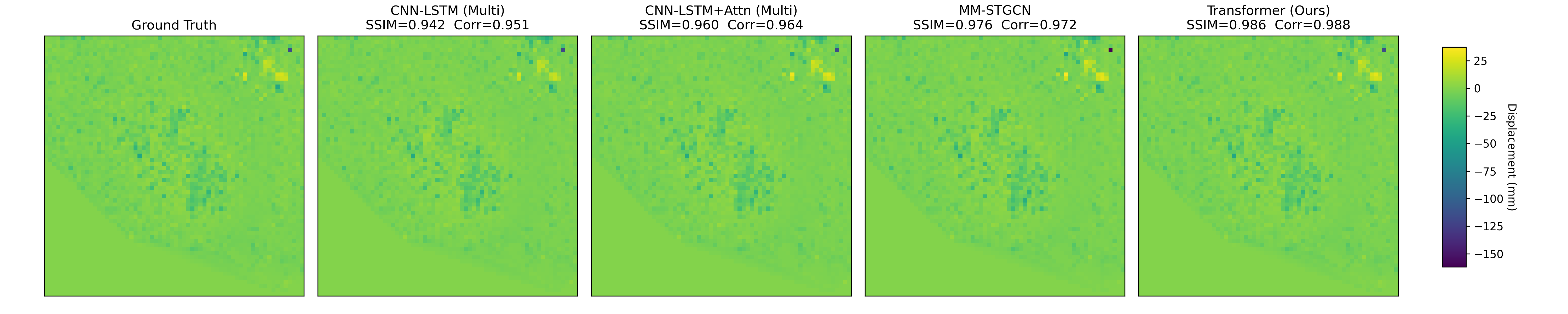}
  \caption{Single–step predicted displacement map comparison for the
  multimodal setting on tile E32N34.}
  \label{fig:map_multimodal}
\end{figure*}

\begin{figure*}[htbp]
  \centering
  \subfigure[In–depth performance analysis for all models in the
  displacement–only setting.]{
    \includegraphics[width=0.48\textwidth]{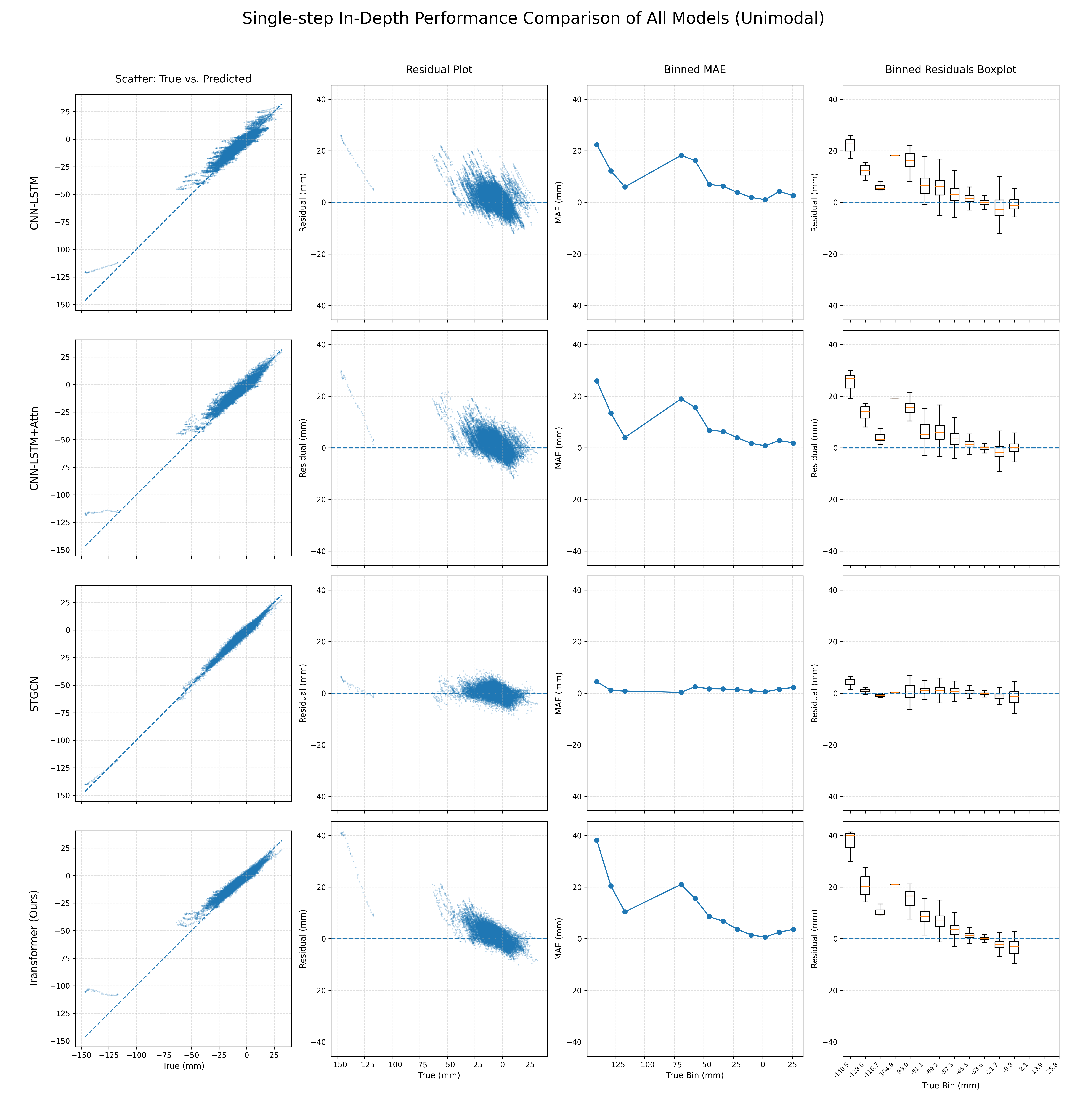}
    \label{fig:analysis_unimodal}
  }
  \hfill
  \subfigure[In–depth performance analysis for all models in the
  multimodal setting.]{
    \includegraphics[width=0.48\textwidth]{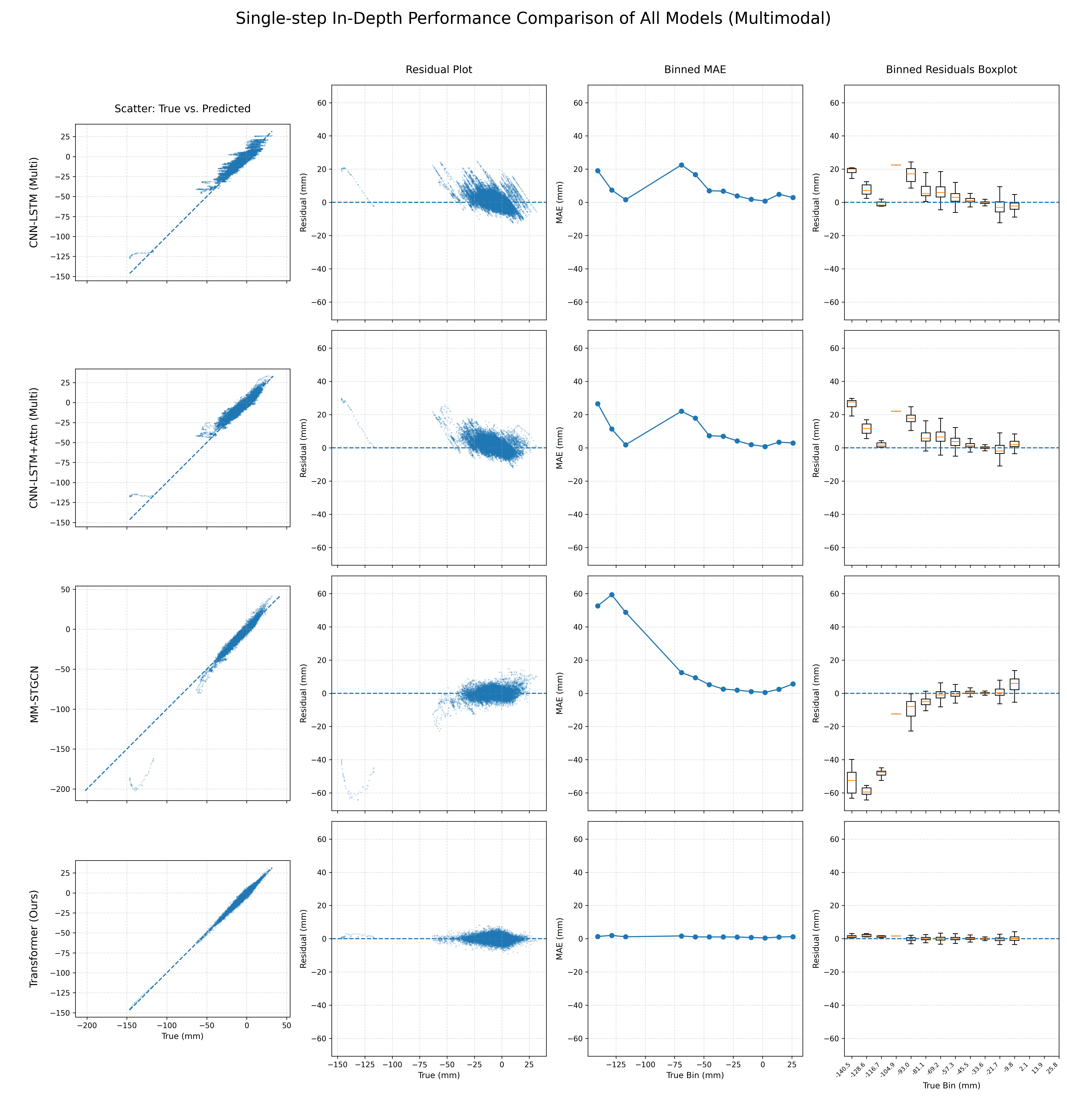}
    \label{fig:analysis_multimodal}
  }
  \caption{Comparison between unimodal and multimodal performance
  analyses on tile E32N34.}
\end{figure*}

\section{Discussion}
\label{sec:discussion}

\subsection{Impact of multimodal conditioning on single–tile forecasting}

The experiments on the eastern Ireland tile (E32N34) demonstrate that
augmenting EGMS displacement time series with static kinematic
descriptors and harmonic time encodings substantially improves the
accuracy and robustness of single–step ground–motion forecasting.
When only displacement maps are used as input
(Section~\ref{sec:experiments}), the STGCN baseline achieves the
best performance among the four architectures with an
$\mathrm{RMSE}=1.04$~mm and $R^{2}=0.96$, while the proposed Transformer
obtains $\mathrm{RMSE}=1.59$~mm and $R^{2}=0.92$. In other words, with
purely kinematic information the graph–based model benefits more from
its strong spatial inductive bias than the more flexible but less
constrained attention architecture.

Once mean velocity, acceleration, seasonal amplitude and harmonic
day–of–year encodings are included
(Section~\ref{sec:experiments}), the ranking is reversed. The
multimodal Transformer reaches $\mathrm{RMSE}=0.90$~mm,
$\mathrm{MAE}=0.57$~mm and $R^{2}=0.97$, clearly outperforming the
multimodal STGCN ($\mathrm{RMSE}=1.45$~mm, $R^{2}=0.93$) and both
CNN--LSTM variants. The recurrent baselines also benefit from the
additional channels, but the absolute gains are modest and do not
fundamentally change their relative position. This shift highlights that
self–attention particularly benefits from rich feature spaces: with
displacement–only input, the Transformer must infer long–term trends,
seasonal cycles and short–term anomalies purely from local temporal
context, whereas with multimodal conditioning these factors are at least
partially disentangled and presented explicitly as separate channels.

\subsection{Threshold accuracies and practical relevance}

Threshold accuracies provide a more operational view of the forecasting
skill, especially for applications that rely on displacement exceedance
criteria. In the displacement–only configuration, the STGCN achieves the
highest $\mathrm{Acc@1mm}$ (about $76\%$), whereas the Transformer
reaches roughly $68\%$, with both models outperforming the CNN–based
baselines across most thresholds. After multimodal conditioning, the
Transformer exhibits the largest relative improvement: $\mathrm{Acc@1mm}$
increases to $81\%$, and the proportion of pixels within $10\%$ relative
error nearly triples compared with the unimodal case
(Section~\ref{sec:experiments}). The multimodal STGCN also
improves, but remains behind the Transformer at the tightest thresholds
(0.5~mm and below).

These gains are particularly relevant for asset-management and
near-real-time monitoring scenarios, where the key question is not only
how small the global RMSE is, but how reliably a model remains within
narrow bands around physically meaningful limits from one acquisition
to the next. In this sense, the proposed single-step predictor is best
viewed as providing short-range updates between consecutive EGMS
epochs, which can complement dedicated early-warning systems that
operate on different time scales and with additional information
(e.g.\ hydrological or engineering data).

\subsection{Error structure and deformation regimes}

The map–based comparisons in Figures~\ref{fig:map_unimodal} and
\ref{fig:map_multimodal} show that all four architectures reproduce the
broad deformation pattern over eastern Ireland, but differ in structural
detail. In the displacement–only case, STGCN and the Transformer achieve
the highest SSIM and pixel–wise correlation with the reference EGMS
map, indicating that both capture the morphology of subsidence bowls and
stable areas reasonably well. However, the deep–analysis panel for the
unimodal setting (Figure~\ref{fig:analysis_unimodal}) reveals clear
magnitude–dependent biases: CNN–based models tend to under–estimate
large subsidence, while the Transformer exhibits more dispersed
residuals at intermediate magnitudes.

In the multimodal configuration, the Transformer delivers the most
faithful reconstruction of fine–scale deformation structures. Its
predicted maps are visually closest to the ground truth, and the
associated SSIM and correlation scores are highest across the four
architectures. The residual analysis in
Figure~\ref{fig:analysis_multimodal} shows that Transformer errors are
tightly concentrated around zero with a nearly flat binned–MAE curve,
suggesting stable accuracy across the full displacement range. The
multimodal STGCN remains competitive, particularly in smoothly varying
regions, but exhibits slightly broader residual distributions and more
skewness in the largest–subsidence bins. These patterns support the
hypothesis that global self–attention is especially effective at
reconciling signals from static and dynamic modalities and at
propagating information across the entire spatial domain.

\subsubsection{Ablation on loss functions}
\label{subsec:loss_ablation}

A potential concern is that the proposed Transformer is trained with a
richer composite loss (Section~\ref{subsec:trainingoptim}) than the
baselines, which might favour it on structural metrics and
threshold-based accuracies. To assess whether our conclusions depend on
this choice, we conduct an ablation in which both the multimodal
Transformer and the multimodal STGCN are retrained on tile E32N34 using
a simple SmoothL1 loss only, i.e.\ without the relative-error,
correlation or gradient terms.

Table~\ref{tab:loss_ablation} summarises the single-step performance
under this setting. Even when trained solely with SmoothL1 loss, the
Transformer clearly outperforms MM--STGCN: RMSE is reduced from
$1.02$~mm to $0.87$~mm, MAE from $0.67$~mm to $0.55$~mm, and $R^{2}$
improves from $0.97$ to $0.97$ (from $0.965$ to $0.974$ in absolute
terms). The proportion of predictions within $1$~mm also increases from
$76.8\%$ to $82.1\%$.

For reference, we also include the Transformer trained with the full
composite loss. As expected, the additional terms further decrease the
error (RMSE $0.74$~mm, MAE $0.41$~mm, $R^{2}=0.98$) and increase
Acc@1mm to $87.4\%$, but they are not required for the Transformer to
surpass MM--STGCN. This ablation confirms that the superior multimodal
performance of the proposed architecture is not an artefact of using a
more complex loss function.

\begin{table}[t]
  \scriptsize
  \centering
  \setlength{\tabcolsep}{1mm}
  \caption{Ablation on loss functions for multimodal single-step
  forecasting on tile E32N34. All models use identical inputs, training
  splits and optimisation settings; only the loss definition differs.}
  \label{tab:loss_ablation}
  \begin{tabular}{lcccc}
    \toprule
    Model \& loss & RMSE [mm] & MAE [mm] & $R^2$ & Acc@1mm [\%] \\
    \midrule
    MM--STGCN (SmoothL1 only)
      & 1.0178 & 0.6684 & 0.9651 & 76.77 \\
    Transformer (SmoothL1 only)
      & 0.8701 & 0.5514 & 0.9737 & 82.12 \\
    Transformer (composite loss)
      & 0.7353 & 0.4092 & 0.9811 & 87.37 \\
    \bottomrule
  \end{tabular}
\end{table}
\subsection{Cross–site behaviour and generalisation across Europe}

The additional experiments on six EGMS tiles representing contrasting
deformation regimes (continuous subsidence, seasonal motion and
co–seismic offsets; Section~\ref{sec:cross-site-results} and
Figure~\ref{fig:maps_generalisation}) provide a first assessment of how
the multimodal Transformer behaves beyond the original training tile.

We note that the set of representative EGMS tiles overlaps with our related study~\citep{yao2025multimodalspatiotemporaltransformerhighresolution},
but the experimental regime differs fundamentally: there models are trained in a site-specific manner,
whereas here we evaluate \emph{zero-shot} transfer from a single source tile (E32N34) without retraining.

When the model is re–trained separately on each tile using the same
architecture and preprocessing, it achieves consistently high
single–step skill across all sites, with $R^{2}$ values typically
between $0.93$ and $0.99$ and $\mathrm{Acc@1mm}$ commonly exceeding
$70\%$ in both continuous and seasonal settings. Even in the challenging
co–seismic tiles affected by the Samos and Croatia earthquakes, the
Transformer maintains high $R^{2}$ and very high
$\mathrm{Acc@50\%}$, indicating that it can capture both the amplitude
and spatial footprint of abrupt offsets once site–specific statistics
are provided.

More importantly for the broader theme of this work, the zero–shot
experiments---where a single multimodal Transformer trained only on
E32N34 is applied without fine–tuning to the other five tiles---show
that much of this skill carries over across Europe. Across all six sites
(including E32N34 itself), the zero–shot model attains
$R^{2}\geq 0.93$, with $\mathrm{Acc@1mm}$ typically in the range of
$70$--$80\%$ for continuous and seasonal regimes and slightly lower but
still competitive values for the co–seismic cases. Tiles dominated by
seasonal or trend–like behaviour (e.g.\ E32N35, E39N30, E44N23) show
particularly strong transfer, whereas large, step–like co–seismic
offsets expose the natural limitations of training on a single,
comparatively smooth reference tile. For the co--seismic tiles, it is important to distinguish between
forecasting an event and being robust to its consequences.
Because our model performs one-step-ahead nowcasting, it does not have access
to physical precursors of an earthquake and cannot predict the co--seismic
jump \emph{before} it occurs.
Instead, its apparent strong performance over co--seismic records is mainly
driven by the post-event period, during which the input window already includes
the new displacement level.
Figure~\ref{fig:coseismic_timeseries} visualises this behaviour: errors peak at
the event epoch and then remain bounded afterwards, illustrating robustness to
sharp discontinuities rather than earthquake prediction.

The additional cross--site experiments with a multimodal STGCN trained
on E32N34 further support this interpretation: although STGCN performs
very well in the site--specific setting, its zero--shot transfer to
other tiles leads to markedly larger errors and much lower threshold
accuracies than those of the transformer. This suggests that the global
self--attention and query--token design of the multimodal transformer
provides a more robust inductive bias for handling regional shifts in
deformation regime and noise characteristics than purely local
spatio--temporal graph convolutions.

Taken together, these results suggest that the proposed multimodal
design does not merely overfit to one specific region, but learns
deformation patterns and modality interactions that are reusable across
sites with different kinematic regimes. At the same time, the gap
between per–tile and zero–shot performance, especially in co–seismic
settings, highlights the value of lightweight fine–tuning or
domain–adaptation strategies when moving to strongly out–of–distribution
areas.

\subsection{Role of architectural choices}

The comparison between CNN--LSTM, CNN--LSTM+Attn, STGCN and the
Transformer clarifies the respective strengths of recurrent, graph and
attention–based designs for InSAR–driven deformation forecasting. The
consistent but moderate improvement of CNN--LSTM+Attn over the plain
CNN--LSTM in all configurations indicates that simple additive temporal
attention helps recurrent networks to focus on informative historical
states, but does not fundamentally change their capacity to model
complex spatio–temporal dependencies.

STGCN excels in the displacement–only setting because it hard–codes the
regular grid topology and performs local message passing over
neighbouring pixels. This inductive bias aligns well with the spatial
continuity of displacement fields and acts as an effective regulariser
when auxiliary inputs are absent. The Transformer, in contrast, treats
each patch as a token and relies on learned positional encodings; its
flexibility is most advantageous when multiple modalities and longer
temporal contexts must be integrated, but can be less data–efficient in
the purely unimodal regime.

The multimodal Transformer configuration adopted here---patch–wise
encoding, temporal query tokens, residual prediction in normalised
space, and a composite loss balancing absolute, relative, structural and
incremental errors—appears to strike a favourable balance between
expressivity and stability. Query tokens decouple the representation of
future states from the input sequence, residual learning focuses the
network on short–term increments rather than absolute levels, and
pixel–wise normalisation ensures that subsidence bowls, seasonal cycles
and co–seismic steps are treated on comparable scales. The strong
multimodal and cross–site performance suggests that these architectural
choices are well suited to dense EGMS–like data.

\subsection{Implications for operational ground–motion monitoring}

From an application perspective, the results indicate that multimodal
learning has practical value for operational deformation monitoring
workflows based on EGMS or similar InSAR products. Within the
constraints of single-step prediction at the Sentinel-1 repeat
interval, the multimodal Transformer can provide map-wide forecasts of
likely displacement at the next acquisition, which is useful for
tracking whether ongoing trends are accelerating, stabilising or
remaining within expected bounds. Static descriptors of historical
displacement (mean velocity, acceleration, seasonal amplitude) can be
computed once from archived time series and re-used as low-cost
covariates, while harmonic time encodings inject seasonal phase
information without requiring external forcing data.

The fact that a single multimodal Transformer trained on one tile
generalises reasonably well across other European sites suggests that a
continental–scale forecasting system could be built around a small
number of pre–trained models, optionally fine–tuned on regions of
special interest. High accuracies at tight thresholds further indicate
that such models could support prioritisation of areas for more frequent
acquisitions, detailed geotechnical inspection or targeted modelling,
particularly where small deviations from background subsidence trends
are operationally significant.

\subsection{Limitations and future work}

Despite these promising results, several limitations remain. First, the
training and primary comparison are still centred on a single EGMS tile
in eastern Ireland, and the cross–site experiments, although covering
different deformation regimes, are restricted to a small set of European
examples. A broader evaluation over a more diverse collection of tiles,
including rapidly subsiding megacities, volcanic regions and
landslide–prone slopes, would provide a more comprehensive picture of
the model’s strengths and failure modes. In this context, systematic
studies of zero–shot transfer, fine–tuning and domain adaptation will be
an important next step.

Second, the present framework focuses on single–step forecasting at a
fixed temporal resolution. Extending the approach to multi–step horizons
and irregular sampling, and integrating it into real–time updating
pipelines, would increase its utility for early–warning and
decision–support systems. A more explicit treatment of predictive
uncertainty---for example via quantile regression, ensembles or
Bayesian variants of the Transformer---would further enhance its
suitability for risk–informed applications.

Finally, the multimodal inputs considered here represent only a subset
of potentially informative covariates. Incorporating additional data
layers such as digital elevation models, land cover, groundwater levels
or anthropogenic loading indicators could both improve forecast skill and
help disentangle different physical drivers of deformation. Combining
the proposed Transformer with interpretability tools (e.g.\ SHAP values,
attention visualisation or gradient–based saliency) will be essential to
understand which modalities, spatial regions and time periods dominate
its predictions, and to bridge the gap between purely data–driven
learning and process–based geophysical understanding.

Overall, the discussion above highlights that multimodal
attention–based models offer a promising and flexible framework for
high–precision land–deformation forecasting from satellite–derived
ground–motion products, while also pointing to clear avenues for further
validation and methodological refinement.

\section{Conclusions and outlook}
\label{sec:conclusion}

This study presented a multimodal deep learning framework for
single–step ground–deformation forecasting from European Ground Motion
Service (EGMS) Level~3 products. Building on a displacement–only
benchmark developed in our previous work, we constructed a unified
six–channel input representation that combines dynamic displacement
history, static descriptors of long–term kinematics and harmonic
time–of–year encodings, and evaluated four representative
spatio–temporal architectures---CNN--LSTM, CNN--LSTM with temporal
attention, a spatio–temporal graph convolutional network (STGCN), and a
patch–based multimodal Transformer---on an EGMS tile covering eastern
Ireland. We then extended the analysis to a set of additional tiles
across Europe representing continuous subsidence, seasonal motion and
co–seismic displacement.

Our main findings can be summarised as follows:

\begin{enumerate}
    \item \textbf{Multimodal conditioning substantially improves predictive skill.}
    Across all architectures, enriching the displacement time series with
    static deformation statistics and harmonic day–of–year information
    yields more accurate and more reliable forecasts than using
    displacement maps alone. The effect is particularly pronounced for
    the Transformer: when moving from the displacement–only to the
    multimodal setting on the eastern Ireland tile, its $\mathrm{RMSE}$
    decreases from about $1.6$~mm to $0.9$~mm and $R^{2}$ increases from
    roughly $0.92$ to $0.97$, with threshold accuracies improving
    systematically from $\mathrm{Acc@10\%}$ down to $\mathrm{Acc@0.1mm}$.

    \item \textbf{A tailored multimodal Transformer outperforms specialised spatio–temporal baselines.}
    In the displacement–only experiments, the STGCN baseline achieves
    the best overall accuracy, reflecting the advantage of exploiting
    explicit grid topology and local message passing when only kinematic
    information is available. In the multimodal configuration, however,
    the proposed Transformer---combining patch–wise embeddings, temporal
    query tokens and residual prediction in normalised space---consistently
    outperforms the multimodal STGCN and both CNN--LSTM variants in
    terms of $\mathrm{RMSE}$, $R^{2}$ and threshold–based accuracy.
    Qualitative comparisons and structural similarity metrics confirm
    that the multimodal Transformer provides the most faithful
    reconstructions of fine–scale deformation patterns.

    \item \textbf{Attention mechanisms are particularly effective in rich feature spaces.}
    The moderate but consistent gain of CNN--LSTM+Attn over the plain
    CNN--LSTM indicates that even simple temporal attention helps
    recurrent models focus on informative historical states. The much
    larger gains achieved by the multimodal Transformer suggest that
    global self–attention is especially powerful when the model must
    fuse heterogeneous spatially distributed inputs (dynamic
    displacement, static summary statistics and temporal encodings) and
    propagate information across the full spatial domain. In this regime,
    attention–based architectures are not only competitive with, but can
    surpass, state–of–the–art spatio–temporal graph models.

    \item \textbf{The multimodal Transformer generalises across European deformation regimes.}
    Additional experiments on six EGMS tiles representing continuous
    subsidence, seasonal ground motion and co–seismic offsets show that
    the proposed architecture maintains high single–step skill well
    beyond the original training tile. When re–trained per site, the
    multimodal Transformer attains $R^{2}$ values typically between
    $0.93$ and $0.99$ with $\mathrm{Acc@1mm}$ often above $70\%$ across
    all regimes. In a stricter zero–shot setting, a single model trained
    only on the eastern Ireland tile still achieves $R^{2}\geq 0.93$ and
    competitive threshold accuracies on the remaining tiles, indicating
    that it learns deformation patterns and modality interactions that
    transfer across diverse European settings.
\end{enumerate}

From an operational perspective, these results indicate that multimodal
attention–based models constitute a promising tool for EGMS–based
deformation monitoring. The additional inputs required in this work
(long–term velocity, acceleration, seasonal amplitude and sinusoidal
time features) can be derived directly from existing InSAR stacks
without external data, making the proposed approach straightforward to
integrate into current processing pipelines. The high accuracies
achieved at strict absolute and relative error thresholds, together with
the encouraging cross–site behaviour, suggest that multimodal
Transformers could support early detection of anomalous ground motion
and prioritisation of areas for further inspection or higher–frequency
acquisitions.

Several avenues remain for future research. First, the present analysis
focuses on single–step forecasting at a fixed temporal resolution; 
extending the framework to multi–step horizons, irregular acquisition
intervals and real–time updating will be important to assess robustness
under more realistic conditions. Second, incorporating additional
covariates such as topography, land cover, groundwater levels or
anthropogenic loading indicators could further enhance predictive skill
and help disentangle different physical drivers of deformation. Third,
combining the proposed multimodal Transformer with uncertainty
quantification and interpretability techniques (e.g.\ ensemble methods,
quantile regression, SHAP analysis or attention visualisation) may help
bridge the gap between purely data–driven forecasts and process–based
understanding of land deformation, thereby increasing the
trustworthiness and uptake of deep learning models in geohazard
assessment and infrastructure management.

\section*{Declaration of generative AI and AI-assisted technologies 
in the manuscript preparation process}

During the preparation of this work the author(s) used ChatGPT 
(OpenAI) and other large language model-based assistants in order to 
improve the clarity and readability of the text, to refine figure 
captions and schematic descriptions, and to obtain suggestions for 
code structuring and response wording. After using these tools, the 
author(s) reviewed and edited all generated content as needed and 
take full responsibility for the content of the published article.

\section*{Acknowledgment}
This research was conducted with the financial support of Science Foundation Ireland under Grant Agreement No.\ 13/RC/2106\_P2 at the ADAPT SFI Research Centre at University College Dublin. ADAPT, the SFI Research Centre for AI-Driven Digital Content Technology is funded by Science Foundation Ireland through the SFI Research Centres Programme. This work is partly supported by China Scholarship Council (202306540013).

\bibliography{./References/longforms,./References/references}


\end{document}